\documentclass[a4paper,UKenglish]{llncs}
\usepackage{tikz}
\usetikzlibrary{positioning, automata, arrows, calc, backgrounds, decorations.pathreplacing}
\usepackage{xcolor}
\usepackage{colortbl}
\usepackage{hyperref}
\usepackage{amsmath}
\usepackage{amssymb}
\usepackage{tcolorbox}
\usepackage{semantic}
\usepackage{subfig}
\usepackage{multirow}
\usepackage{lineno}
\usepackage{mathtools}
\usepackage{xifthen}
\usepackage{empheq}
\usepackage{framed}
\usepackage{wrapfig}

\newcommand{\scaledInference}[3][]{\scalebox{0.8}{\inference[#1]{#2}{#3}}}

\newcommand{\automaton}{\mathcal{M}}
\newcommand{\alphabet}{\Sigma}
\newcommand{\letter}{\sigma}
\newcommand{\transition}{\delta}
\newcommand{\kstar}{^{*}}
\newcommand{\pushdown}{\mathcal{K}}

\newcommand{\newsemantics}[2]{\newcommand{#1}{\mathsf{#2}}}
\newcommand{\newfunction}[2]{\newcommand{#1}{\mathop\mathrm{#2}}} %
\newcommand{\newcomponent}[2]{\newcommand{#1}{\mathsf{#2}}}
\newcommand{\newaction}[2]{\newcommand{#1}{\mathtt{#2}}}
\newcommand{\renewaction}[2]{\renewcommand{#1}{\mathtt{#2}}}
\newcommand{\newadt}[2]{\newcommand{#1}{\textsc{#2}}}

\newcommand{\sizeof}[1]{|#1|}
\newcommand{\lengthof}[1]{|#1|}
\newcommand{\of}[1]{(#1)}
\newcommand{\tuple}[1]{\langle#1\rangle}
\newcommand{\set}[1]{\{#1\}}

\newcommand{\cof}[1][]{\ifthenelse{\isempty{#1}}{}{\of{#1}}}
\renewcommand{\to}[1][]{\mathop{\xrightarrow{~#1~}}}
\newcommand{\word}{w}

\newcommand{\newclass}[2]{\newcommand{#1}{\textsc{#2}}}
\newclass{\exptime}{ExpTime}
\newclass{\etime}{ETime}
\newclass{\nexptime}{NExpTime}
\newclass{\class}{Class}
\newclass{\pspace}{PSpace}
\newclass{\expspace}{ExpSpace}
\newclass{\dtime}{DTime}
\newclass{\dspace}{DSpace}

\newcommand{\TS}{\mathcal{T}}

\newcomponent{\conf}{c}
\newcomponent{\confset}{C}
\newcomponent{\lbl}{lab}
\newcomponent{\lblset}{Labs}
\newcomponent{\state}{q}
\newcomponent{\stateset}{Q}

\newcommand{\step}[2][]{\to[#1]_{#2}}
\newcommand{\upwardclosure}{\mathop\uparrow}
\newfunction{\pos}{pos}
\newfunction{\pre}{pre}

\newsemantics{\init}{init}
\newsemantics{\final}{final}
\newsemantics{\target}{target}
\newsemantics{\all}{all}
\newsemantics{\diff}{diff}

\newcommand{\adt}{\mathsf{A}}
\newadt{\noadt}{NoAdt}
\newadt{\petri}{Petri}
\newcomponent{\val}{val}
\newcomponent{\vals}{Vals}
\newcomponent{\op}{op}
\newcomponent{\ops}{Ops}
\newcommand{\adtstep}[2]{\step[#1]{#2}}

\newcommand{\stackalphabet}{\Gamma}
\newcommand{\stacksymbol}{\gamma}
\newcommand{\stackword}{w}
\newcommand{\emptystack}{\varepsilon}
\newadt{\stack}{St}
\newaction{\pop}{pop}
\newaction{\push}{push}
\newaction{\isEmpty}{isEmpty}
\newadt{\multistack}{OMSt}

\newadt{\Counter}{Ct}
\newcommand{\counter}[1][]{\Counter\cof[#1]}

\newadt{\WkCounter}{wCt}
\newcommand{\wkcounter}[1][]{\WkCounter\cof[#1]}

\newaction{\inc}{inc}
\newaction{\dec}{dec}
\newaction{\isZero}{isZero}
\newcommand{\nof}[1]{n\text{-}#1}
\newcommand{\hoadt}[2]{{#1} \text{-} #2}
\renewcommand{\program}{\mathcal{P}}
\newcomponent{\process}{Proc}
\newcomponent{\tso}{TSO}
\newcomponent{\dtso}{DTSO}
\newcommand{\run}{\rho}
\newcomponent{\plbl}{plab}

\newcomponent{\xvar}{x}
\newcomponent{\varset}{Vars}
\newcomponent{\dval}{d}
\newcomponent{\valset}{Dom}
\newcomponent{\msg}{msg}
\newcomponent{\msgs}{Msgs}
\newcomponent{\instr}{instr}
\newcomponent{\instrs}{Instrs}
\newcommand{\xd}{\xvar, \dval}

\newaction{\rd}{rd}
\renewaction{\wr}{wr}
\renewaction{\skip}{skip}
\newaction{\mf}{mf}
\newaction{\up}{up} %

\newcommand{\indexset}{\mathcal{I}}
\newcommand{\statemap}{\mathcal{S}}
\newcommand{\valuemap}{\mathcal{V}}
\newcommand{\buffermap}{\mathcal{B}}
\newcommand{\memorymap}{\mathcal{M}}
\newcommand{\pid}{\iota}
\newcommand{\buffer}{\beta}
\newcommand{\emptybuffer}{\varepsilon}

\newfunction{\pivot}{pvt}
\newfunction{\lval}{LVal}
\newfunction{\rval}{ReadVal}

\newcomponent{\view}{v}
\newcomponent{\viewset}{V}
\newcomponent{\bookkeepset}{B}
\newcomponent{\lastwrite}{Lw}
\newcomponent{\rank}{rk}
\newcomponent{\ptr}{\phi}
\newcommand{\eptr}{\phi_E}
\newcommand{\lptr}{\phi_L}
\newcommand{\pptr}{\phi_P}
\newcommand{\maxlptr}{\lptr^{\max}}

\newcommand{\upseq}{\omega}
\newcommand\none\oslash
\newcommand{\pvt}{\mathsf{pvt}}

\newcommand{\RM}{\mathcal{R}\of\adt}
\newcommand{\RMof}[1]{\mathcal{R}\of{#1}}
\newcomponent{\reg}{reg}
\newcomponent{\rg}{r}
\newcommand{\rgd}{\rg, \dval}
\newcomponent{\regs}{Regs}
\newcomponent{\act}{act}
\newcomponent{\acts}{Acts}

\newaction{\SKP}{SKP}
\newaction{\INC}{INC}
\newaction{\DEC}{DEC}
\newaction{\CKZ}{CKZ}
\newaction{\CKNZ}{CKNZ}
\newaction{\CPY}{CPY}
\newaction{\CMPL}{CMPL}
\newaction{\WRT}{CMPL}
\newaction{\CKE}{CKE}
\newaction{\CKNE}{CKNE}
\newaction{\CKL}{CKL}
\newaction{\CKG}{CKG}
\newaction{\CKLE}{CKLE}
\newaction{\CKGE}{CKGE}
\newaction{\SET}{SET}
\newaction{\READ}{READ}
\newaction{\WRITE}{WRITE}

\newcommand{\tsopreach}{\texttt{$\tso\of\adt$-P-Reach}}
\newcommand{\tsopreachof}[1]{\texttt{$\tso\of{#1}$-P-Reach}}

\newcommand{\rmreach}{\texttt{$\RM$-Reach}}
\newcommand{\rmreachof}[1]{\texttt{$\RMof{#1}$-Reach}}
\newcommand{\fsmreach}{\texttt{$\automaton\of{\adt}$-Reach}}
\newcommand{\fsmreachof}[1]{\texttt{$\automaton\of{#1}$-Reach}}

\newcommand{\problembox}[3]{
	\begin{framed}
		\noindent\textbf{#1:}

		\noindent\textbf{Given:}#2

		\noindent\textbf{Decide:}#3
	\end{framed}
}

\begin{document}
\title{Parameterized Verification under TSO with Data Types}

\institute{Uppsala University, Sweden \and
	Indian Institute of Technology Bombay
	\and UC Berkeley, USA
}
\author{Parosh Abdulla\inst{1} \and
	Mohamad Faouzi Atig	\inst{1} \and
	Florian Furbach\inst{1}  \and
	Adwait Godbole  \inst{3} \and
	Yacoub G. Hendi\inst{1} \and
	Shankaranarayanan Krishna	\inst{2} \and
	Stephan Spengler\inst{1}}
\authorrunning{Parosh A. Abdulla et al.}

\maketitle
We consider parameterized verification of systems executing according to the total store ordering (TSO) semantics.
The processes manipulate abstract data types over potentially infinite domains.
We present a framework that translates the reachability problem for such systems to the reachability problem for register machines enriched with the given abstract data type.
We use the translation to obtain tight complexity bounds for TSO-based parameterized verification over several abstract data types, such as push-down automata, ordered multi push-down automata, one-counter nets, one-counter automata, and Petri nets. We apply the framework to get complexity bounds for higher order stack and counter variants as well.

\section{Introduction}
A \emph{parameterized system} consists of a fixed but arbitrary number of
identical processes that execute in parallel.
The goal of \emph{parameterized verification} is to prove the correctness
of the system regardless of the number of processes.
Examples for such systems are sensor networks, leader election protocols, and mutual exclusion protocols.
The topic has been the subject of intensive research
for more than three decades (see e.g. \cite{DBLP:journals/ipl/AptK86,DBLP:journals/jacm/GermanS92,DBLP:journals/sigact/BloemJKKRVW16,DBLP:journals/sttt/AbdullaD16}), and it is the subject of one chapter of the
Handbook of Model Checking \cite{DBLP:reference/mc/AbdullaST18}.
Research on parameterized verification has been mostly conducted under the premise that
(i) the processes run according to the classical
Sequential Consistency (SC) semantics, and
(ii) the processes are finite-state machines.

Under SC, the processes operate on a set of shared variables through which they communicate \emph{atomically}, i.e., read and write operations take effect immediately.
In particular, a write operation is visible to all the processes as soon as the writing process carries out its operation.
Therefore, the processes always maintain a uniform view of the shared memory: they all see the latest value written on any given variable, hence
we can interpret program runs as interleavings of sequential process executions.
Although SC has been immensely popular as an intuitive way of understanding the behaviours of concurrent processes, it is not realistic to assume computation platforms guarantee SC anymore.
The reason is that, due to hardware and compiler optimizations, most modern platforms allow more relaxed program behaviours than those permitted under SC,
leading to so-called \emph{weak memory models}.
Weakly consistent platforms are found at all levels of system design
such as multiprocessor architectures
(e.g., \cite{SSONM2010,SarkarSAMW11}),
Cache protocols (e.g., \cite{aros-micro16,ElverN14}),
language level concurrency (e.g., \cite{LahavGV16}), and
distributed data stores (e.g., \cite{Burckhardt14}).
Therefore, in recent years, research on the
parameterized verification of concurrent programs under weak memory models
have started to become popular.
Notable examples are the cases of the TSO semantics \cite{par_atig} and the Release-Acquire semantics of C11 \cite{DBLP:conf/podc/KrishnaG0C22}.

In a parallel development, several works have extended the basic model of parameterized systems (under the SC semantics) by considering processes that are infinite-state systems.
The most dominant such class has been the case where the individual processes are variants of push-down automata \cite{DBLP:conf/lics/Kahlon08,DBLP:conf/fsttcs/Hague11,DBLP:journals/jacm/EsparzaGM16,DBLP:journals/jacm/EsparzaGM16,DBLP:conf/concur/TorreMW15,DBLP:conf/vmcai/MuschollSW17,DBLP:conf/cav/FortinMW17}

Parameterized verification is difficult, even under the original assumption of both SC and finite-state processes as we still need to handle an infinite state space. %
The extension to weakly consistent systems is even more complex due to the intricate extra process behaviours.
Almost all weak memory models induce infinite state spaces even without parameterization and even when the program itself is finite-state.
Therefore, performing parameterized verification under weak consistency requires handling a state space that is infinite in two dimensions; one due to parameterization and one due to the weak memory model.
The same applies to the extension of parameterized verification under SC where the processes are infinite-state:
in addition to infiniteness due to parameterization, we have a second source of infinity due to the infiniteness of the processes.

In this paper, we combine the above two extensions.
We study parameterized verification of programs under the TSO semantics, where the processes use infinite data structures such as stacks and counters.
The framework is uniform in that the manipulation can be described using an abstract data type.

We revisit the pivot abstraction technique presented in \cite{par_atig}.
As a first contribution, we show that we can capture pivot abstraction precisely, using a class of register machines in which the registers assume values over a finite domain.
We show that, for any given abstract data type $\adt$, we can reduce, in polynomial time, the parameterized verification problem under TSO and $\adt$ to the reachability problem for register machines manipulating $\adt$.
Furthermore, we show that the reduction also holds in the other direction: the reachability problem for register machines over $\adt$ is polynomial-time reducible to the parameterized verification problem under TSO for $\adt$.
In particular, the model abstracts away the semantics of TSO (in fact, it abstracts away concurrency altogether) since we are dealing with a single register machine.

We summarize the contributions of the paper as follows:
\begin{itemize}
\item
We present a register abstraction scheme that captures the behaviour of parameterized systems under the TSO semantics.
\item
We translate  parameterized verification under the TSO semantics when the processes manipulate an ADT $\adt$, to the reachability problem for register machines operating over $\adt$.
\item
We instantiate the framework for deciding the complexity of parameterized verification under TSO for different abstract data types.
In particular we show the problem is \pspace-complete when $\adt$ is a one-counter, $\exptime$-complete if $\adt$ is a stack, $\hoadt{2}{\etime}$-complete if $\adt$ is an ordered multi stack, and \expspace-complete if $\adt$ is a Petri net. We obtain further complexity bounds for higher order counter and stacks.
\end{itemize}

\paragraph{Related Work}
There has been an extensive research effort on parameterized verification since the 1980s
(see \cite{DBLP:journals/sigact/BloemJKKRVW16,DBLP:reference/mc/AbdullaST18} for recent surveys of the field).
Early works showed the undecidability of the general problem (even assuming finite-state processes) \cite{DBLP:journals/ipl/AptK86}, and hence the emphasis has been on finding useful special cases.
Such cases are characterized by three aspects, namely the system topology
(un-ordered, arrays, trees, graphs, rings, etc.),
the allowed communication patterns
(shared memory, Rendez-vous, broadcast, lossy channels, etc.),
and the process types (anonymous, with IDs, with priorities, etc.)
\cite{DBLP:conf/lics/EsparzaFM99,DBLP:conf/concur/DelzannoSZ10,DBLP:journals/toplas/GantyM12,DBLP:conf/csl/EmersonK04,DBLP:conf/charme/EmersonK03,DBLP:conf/tacas/NamjoshiT16}.

Another line of research to counter undecidability are over-approximations based on regular model checking
\cite{DBLP:journals/tcs/KestenMMPS01,DBLP:conf/cav/BoigelotLW03,DBLP:journals/sttt/BouajjaniHRV12,DBLP:journals/sttt/Abdulla12},
monotonic abstraction
\cite{DBLP:conf/concur/AbdullaCDHHR10}, and
symmetry reduction
\cite{DBLP:conf/cav/KaiserKW10,DBLP:conf/lics/EmersonHT00,DBLP:journals/sttt/AbdullaHH16}.

A seminal work in the area is the paper by German and Sistla
\cite{DBLP:journals/jacm/GermanS92}.
The authors consider the verification of systems
consisting of an arbitrary number of finite-state processes
interacting through Rendez-Vous communication.
The paper shows that the model checking problem is \expspace-complete.
In a series of more recent papers, parameterized verification
has been considered in the case where the individual processes
are push-down automata.
\cite{DBLP:conf/lics/Kahlon08,DBLP:conf/fsttcs/Hague11,DBLP:journals/jacm/EsparzaGM16,DBLP:conf/concur/TorreMW15,DBLP:conf/vmcai/MuschollSW17,DBLP:conf/cav/FortinMW17}.%
All the above works assume the SC semantics.

Due to the relevance of weak memory models in parameterized verification, papers on the topic have started to appear in the last two years.
The paper \cite{par_atig} considers
parameterized verification of programs running
under TSO, and shows that the reachability problem is \pspace-complete.
However, the paper assumes that the processes are finite-state and, in particular, the processes do not manipulate unbounded data domains.
The model of the paper corresponds to the particular case of our framework where we take the abstract data type to be empty.
In this case our framework also implies \pspace-completeness.

The paper \cite{DBLP:conf/podc/KrishnaG0C22} shows \pspace-completeness when the underlying semantics is the Release-Acquire fragment of C11.
The latter semantics gives rise to different semantics compared to TSO.
The paper also considers finite-state processes.

The paper \cite{DBLP:journals/lmcs/AbdullaABN18}
considers parameterized verification of programs running
under TSO.
However, the paper applies the framework of well-structured systems
where the buffers of the processes are modeled as lossy channels,
and hence
the complexity of the algorithm is non-primitive recursive.
In particular, the paper does not give
any complexity bounds for the reachability
problem (or any other verification problems).
Conchon et al. \cite{10.1007/s10817-020-09565-w} address the parameterized verification of programs under TSO as well. They make use of Model Checker Modulo Theories, no decidability or complexity results are given.
The paper \cite{DBLP:conf/esop/BouajjaniDM13} considers checking the robustness
property against SC for parameterized systems running
under the TSO semantics.
However, the robustness problem is entirely different
from reachability and the techniques and results developed in the paper
cannot be applied in our setting.
The paper shows that the problem is {\expspace}-hard.
All these works assume finite-state processes.

In contrast to all the above works, the current paper is the first paper that studies decidability and complexity of parameterized verification under the TSO semantics when the individual processes are infinite-state.

\section{Preliminaries}
We denote a function $f$ between sets $A$ and $B$ by $f: A \to B$.
We write $f[a \leftarrow b]$ to denote the function $f'$ such that $f'(a) = b$ and $f'(x) = f(x)$ for all $x \neq a$.

For a finite set $A$, we use $\sizeof{A}$ to refer to the size of $A$.
We also use $A^{*}$ to denote the set of words over $A$ including the empty word $\epsilon$.
For a word $w \in A^{*}$, we use $\lengthof{w}$ to refer to the length of $w$.
We say a word $w$ is \emph{differentiated} if all symbols in $w$ are pairwise different. %
The set $A^\diff$ is the set of all differentiated words over the set $A$.
Finally, for a differentiated word $w$, we define $\pos(w)(a) $ as the unique position of the letter $a$ in $w$.

A \emph{labelled transition system} is a tuple $\tuple{\confset, \confset_\init, \lblset, \to}$, where $\confset$ is the set of configurations, $\confset_\init \subseteq \confset$ is the set of initial configurations, $\lblset$ is a finite set of labels and $\to \subseteq \confset \times \lblset \times \confset$ is the transition relation over the set of configurations.
For a transition $\tuple{\conf_1, \lbl, \conf_2} \in \to$, we usually write $\conf_1 \to[\lbl] \conf_2$ instead.
We use $\conf_1 \to \conf_2$ to denote that $\conf_1 \to[\lbl] \conf_2$ for some $\lbl \in \lblset$.
Furthermore, we write $\to[*]$ to denote the transitive reflexive closure over $\to$, and if $\conf_1 \to[*] \conf_2$ %
then we say $\conf_2$ %
is \emph{reachable from} $\conf_1$. %
If $\conf_1 \in \confset_\init$, then we just say that $\conf_2$ is \emph{reachable}.
A \emph{run} $\run$ is an alternating sequence of configurations and labels and is expressed as follows:
$ \conf_0 \to[\lbl_1] \conf_1 \to[\lbl_2] \conf_2 \dots \conf_{n-1} \to[\lbl_n] \conf_n ~.$
Given $\run$, we write $\conf_0 \to[n] \conf_n$ meaning that $\conf_n$ is reachable from $\conf_0$ by $n$ steps, and we write $\conf_0 \to[\rho] \conf_n$ meaning that $\conf_n$ is reachable from $\conf_0$ through the run $\run$.
\section{Abstract Data Types (ADT)}
\label{sec:adt}
In this section, we introduce the notion of abstract data types (ADTs) which will be used extensively in the paper.
An ADT is a labelled transition system $\adt = \tuple{\vals, \set{\val_\init}, \ops, \step\adt}$. Intuitively, this describes the behaviour of some data type such as a stack, or a counter. $\vals$ is the set of configurations of $\adt$. It describes the possible values the data type can assume. The initial configuration is $\val_\init \in \vals$. The set of labels $\ops$ represents the operations that can be executed on the data type and the transition relation $\step\adt \in \vals \times \ops \times \vals$ describes the semantics of these operations. Below, we give some concrete examples of abstract data types.

\begin{example}[Counter]\label{ex:1counter}
	We define a counter, denoted by the ADT $\Counter$, as follows.
	The set of configurations $\vals^\Counter = \mathbb{N}$ are the natural numbers.
	The initial value, denoted by $\val_\init^\Counter$, is $0$.
	The set of operations is $\ops^\Counter = \set{ \inc, \dec, \isZero }$.
	The transition relation $\step\Counter$ is as follows: %
	The operations $\inc$ and $\dec$ increase or decrease the value of the counter by one, respectively.
	The latter operation is only enabled if the value of the counter is non-zero, otherwise it blocks.
	Finally, the transition $\isZero$ checks that the value of the counter is zero, i.e. it is only enabled if that condition is true.

\end{example}

\begin{example}[Weak Counter]\label{ex:weak_1counter}
	A weak counter differs from a counter in that it cannot be checked for zero.
	The ADT $\WkCounter$ representing a weak counter is defined as in \autoref{ex:1counter}, except the operations of $\WkCounter$ are reduced to $\ops^\WkCounter = \set{ \inc, \dec}$.
\end{example}

\begin{example}[Stack]\label{ex:stack}
Let $\stackalphabet$ be a finite set representing the stack alphabet.
A \emph{stack} $\stack = \tuple{\vals^\stack, \{\val_\init^\stack\}, \ops^\stack, \step\stack}$ on $\stackalphabet$ is defined as follows.
The configurations of $\stack$ are $\vals^\stack = \stackalphabet\kstar$ and the initial configuration is the empty stack $\val_\init^\stack = \emptystack$.
The set of operations is $\ops^\stack = \set{ \pop\of\stacksymbol, \push\of\stacksymbol, \isEmpty \mid \stacksymbol \in \stackalphabet }$.
The transition relation  is as follows.
For every word $\word \in \stackalphabet\kstar$ and every symbol $\stacksymbol \in \stackalphabet$, $\push\of\stacksymbol$ adds the symbol $\stacksymbol$ to the top of the stack.
Similiarly, the $\pop\of\stacksymbol$ operation removes the topmost symbol from the stack.
It is only enabled  if
the topmost symbol on the stack.
The $\isEmpty$ operation does not change the stack, but can only be performed if the stack is the empty word $\emptystack$.
\end{example}

\begin{example}[Petri Nets]\label{ex:PN}
Given a Petri net\cite{Peterson:1981:PNT:539513}, We can define a corresponding ADT $\petri$ that models its semantics.
The values are the markings, the operations are the Petri net transitions and the transition relation is given by the input and output vectors of the Petri net transitions.
\end{example}

\paragraph{Higher Order ADTs}
We extend the ADT $\stack$ to higher order stacks referred to as $\nof{\stack} $.
This is done recursively\cite{DBLP:journals/corr/abs-0705-0262,ENGELFRIET199121}. The formal definition is in Appendix~\ref{app:ho_stack}.
A value of a level $n$ higher order stack $\nof{\stack}$ is a stack of level $n-1$ stacks.
For level 1, it is the standard stack $\stack$. The operations for level $n$ are
$\ops^{\hoadt{n}{\stack}} = \{\pop(\stacksymbol), \push(\stacksymbol), \pop_k, \push_k, ~  | ~ \stacksymbol \in \stackalphabet, 2 \leq k \leq n\}$. The operations $\pop(\stacksymbol)$ and $ \push(\stacksymbol)$ are recursively applied to the top element in the stack (which consists of a stack that is one level lower) until the level of the top element is 1. Here, they have the standard stack behaviour.
Operations $\pop_k$ and $ \push_k$ are recursively applied to the top element until the level of the element is $k$. Then, that level $k$ stack is copied and the copy is pushed on top of the original.

Since a counter can be seen as a stack with an alphabet of size $1$ (and a bottom element $\bot$), we can extend definitions of $\wkcounter$ and $\counter$ to $\nof{\wkcounter}$ and $\nof{\counter}$ in the same way. We add operations $\inc_k, \dec_k$.
All operations are recursively applied to the top counter.
For $\inc, \dec, \isZero$, we use standard behaviour once the level is 1.
For $\inc_k, \dec_k$, we copy/remove the top element once the level is $k$.

\begin{example}[Ordered Multi Stack]
	We extend the stack to a numbered list of n many stacks \nof{\multistack} \cite{10.1007/978-3-540-85780-8_9}.
	A value of \nof{\multistack} consists of list of stacks $\val_1^\stack \ldots \val_n^\stack$. An operation $\ops^{\nof{\multistack}}=\{\isZero_i, \pop_i(\stacksymbol), \push_i(\stacksymbol), ~  | ~ \stacksymbol \in \stackalphabet, i \leq n\}$ works on stack number $i$ in the standard way. One additional condition is that the stacks have to be ordered, meaning an operation $\pop_i(\stacksymbol)$ is only enabled if the stacks $1\ldots i-1$ are empty.
\end{example}
\newpage

\section{TSO with an Abstract Data Type : $\tso\of{\adt}$}

In this section, we introduce concurrent programs running under $\tso\of{\adt}$ for an ADT $\adt = \tuple{ \vals, \set{ \val_\init }, \ops, \step\adt }$.
These programs consist of concurrent processes where the communication between processes is performed using shared memory under the TSO semantics.
In addition, each process maintains a local variable of type $\adt$.

\noindent \textbf{Syntax of $\tso\of{\adt}$}.
Let $\valset$ be a finite data domain and $\varset$ be a finite set of shared variables over $\valset$.
Let $\dval_\init \in \valset$ be the initial value of the variables.
We define the \emph{instruction set} of $\tso\of{\adt}$ as $\instrs = \set{ \rd\of\xd, \wr\of\xd \mid \xvar \in \varset, \dval \in \valset } \cup \set{ \skip, \mf }$,
which are called \emph{read}, \emph{write}, \emph{skip} and \emph{memory fence}, respectively.

A process is represented by a finite state transition system.
It is given by the tuple $\process = \tuple{ \stateset, \state_\init, \transition }$, where $\stateset$ is a finite set of states, $\state_\init \in \stateset$ is the initial state, and $\transition \subseteq \stateset \times (\instrs \cup \ops) \times \stateset$ is the transition relation.
We call this tuple the \emph{description} of the process.
A concurrent program is a tuple of processes $\program = \tuple{\process_\pid}_{\pid \in \indexset}$, where $\indexset$ is some finite set of process identifiers.
For each $\pid \in \indexset$ we have $\process^\pid =  \tuple{ \stateset^\pid, \state_\init^\pid, \transition^\pid }$.

\noindent \textbf{Semantics of $\tso\of{\adt}$}.
We describe the semantics of a program $\program$ running under $\tso\of\adt$ by a labelled transition system $\TS_\program = \tuple{ \confset^\program, \confset_\init^\program, \lblset^\program, \step\program }$. The formal definition is given in Appendix~\ref{sec:app:tso-semantics}.
Under $\tso\of{\adt}$, there is an unbounded FIFO buffer of writes between each process and the memory.
A configuration $\conf\in\confset^\program$ of the system consists of the value of each variable in the shared memory as well as for each process: its local state, its value of the ADT, and the content of the corresponding write buffer.

The labelled transitions $\step\program$ are as follows:
A local $\skip$ transition simply updates the state of the corresponding process.
An ADT operation additionally updates the ADT value according to ADT behaviour $\step\adt$.
When a process executes a write instruction, the operation is enqueued as a \emph{pending write message} into its buffer.
A message $\msg$ is an assignment of the form $\msg=\tuple\xd$, where $\xvar \in \varset$ and $\dval \in \valset$.
We denote the set of all messages by $\msgs = \varset \times \valset$.
The buffer content for a process is given as a word over $\msgs$.
The messages inside each buffer are moved non-deterministically to the main memory in a FIFO manner.
Once a message reaches the memory, it becomes visible to all the other processes.
When executing a read instruction on a variable $\xvar \in \varset$, the process first checks its buffer for pending write messages on $\xvar$.
If the buffer contains such a message, then it reads the value of the most recent one.
If the buffer contains no write messages on $\xvar$, then the process fetches the value of $\xvar$ from the memory.
The \emph{initial configuration} is $\conf_\init^\program $, where each process is in its initial state, each ADT holds its initial value, each store buffer is empty and the memory holds the initial values of all variables.
Note that since FIFO buffer is unbounded, this is an infinite state transition system, even for finite ADT.

A sequence of transitions $\conf_0 \step[\lbl_1]\program \conf_1 \step[\lbl_2]\program \conf_2 \dots \conf_{n-1} \step[\lbl_n]\program \conf_n$ where $\conf_0 = \conf_\init^\program$ is the initial configuration and $\lbl_i\in \lblset^\program$
 is called a run in the $\tso\of{\adt}$ transition system. If there is a run ending in a configuration with state $\state_\final$, then we say $\state_\final$ is reachable by $\process$ under $\tso\of{\adt}$.

\section{Parameterized Reachability in $\tso\of{\adt}$}

In this section, we consider the parameterized TSO setting which allows for an a priori unbounded number of processes with the same process description.
We begin by formally introducing the parameterized state reachability problem, and then develop a generic construction that allows us to represent the TSO semantics (except for the ADT) in a finite manner.

\paragraph*{The Parameterized State Reachability Problem}
Intuitively, parameterization allows for an arbitrary number of identical processes.
The parameterized state reachability problem for $\tso\of{\adt}$ called \tsopreach\ identifies a family of (standard) reachability problem instances.
 We want to determine whether we have reachability in some member of the family.
We now introduce this formally.

For a given process description $\process$, we consider the program instance, $\program_\process^n$ parameterized by a natural number $n$ as follows.
For $\indexset = \{1, \hdots, n\}$, let $\program_\process^n = \tuple{ \process_1, \hdots, \process_n }$ with $\process_\pid = \process$ for all $\pid \in \indexset$.
That is, the $n^\text{th}$ \textit{slice} of the parameterized family of programs contains $n$ processes, all with identical descriptions $\process$.
We require that all processes maintain copies of the ADT $\adt$.

\problembox{
    \tsopreach
}{ %
    A process $\process = \tuple{ \stateset, \state_\init, \transition }$, an ADT $\adt$, and a state $\state_\final \in \stateset$,
}{ %
    Is there a  $n \in \mathbb{N}$  s.t. $\state_\final$ is reachable by $\program_\process^n$ under $\tso\of{\adt}$?
}

When talking about a certain family of ADTs, e.g. the family of petri nets, we write $\tsopreachof{\petri}$ and mean the restriction of \tsopreach\ to petri nets, i.e. to instances where $\adt$ is a petri net.

The main difference between the non-parameterized case and the parameterized case of the problem is that in the first case the index set $\indexset$ is \emph{a priori} fixed, while in the second case it can be \emph{arbitrary}.
This results in  $\confset_\init^\program$ being a singleton in the non-parameterized case while it becomes infinite (one initial state for each $n$-slice) in the parameterized case.

We determine upper and lower bounds for the complexity of the state reachability problem.
The challenge of solving this problem varies with the ADT.
This problem for plain TSO without an ADT has been studied in \cite{par_atig}.
They showed that the problem can be decided in \pspace\ and is in fact \pspace-complete.
The result is based on an abstraction technique called the \textit{pivot} semantics.
The pivot semantics is \emph{exact} in the sense that a state $\state$ is reachable under parameterized TSO if and only if it is reachable under the pivot semantics.

We show that the dynamics underlying the pivot abstraction can be generalized to our model with ADT.
We show that the pivot abstraction can be extended to obtain a register machine. %
We use this construction to give a general characterization of \tsopreach.
First, we recall the pivot abstraction.

\noindent\textbf{The Pivot Abstraction}~\cite{par_atig}.
For a set of variables $\varset$ and data domain $\valset$, processes generate pending write messages from the set $\msgs = \varset \times \valset$ by executing $\wr$ instructions.
This set has size $\sizeof{\varset} \cdot \sizeof{\valset}$ and hence at most as many distinct (variable, value) pairs can be produced in any run.
For a run $\run$ of the program, for each message $\msg = \tuple\xd \in \msgs$ we can define the first point along $\run$ at which some write on variable $\xvar$ with value $\dval$ is propagated to the memory.
The pivot abstraction identifies these points as \emph{pivot} points $\pivot\of\msg$, for each distinct message in $\msgs$.
For a write message $\msg$ under $\run$, the pivot point $\pivot\of{\msg}$ is the first point of propagation of $\msg$ to the memory under $\run$.

The core observation is that if at some point in $\run$, a process $\process_\pid$ propagates a message $\msg = \tuple\xd$ from its buffer to the memory, then after that point, the value $\dval$ will always be available to read on variable $\xvar$ from the shared memory.
Technically, this follows from parameterization.
There are arbitrarily many processes executing identical descriptions.
This means transitions of the original process $\process_\pid$ can be mimicked by a \emph{clone} process $\process_{\pid'}$ identical to $\process_\pid$.
Hence, $\process_{\pid'}$ can replicate the execution of $\process_\pid$ right up to the point where the message $\msg$ is the oldest message in its buffer.
Then a single propagate step updates the value of $\xvar$ in the shared memory to $\dval$.
There can be arbitrarily many such clones and the propagate step can happen at any time.
It follows that beyond the $\pivot\of\msg$ point in $\run$, the value $\dval$ can always be read from $\xvar$.

For distinct messages from $\msgs$, we can order the pivot points corresponding to these messages according to the order in which they appear in $\run$.
This gives us a \emph{first update sequence}, denoted by $\upseq$.
No two messages in $\upseq$ are the same; the set of such sequences is the set of differentiated words $\msgs_\diff$.
A message $\msg \in \msgs$ in $\upseq$ has the \emph{rank} $k$ if it is the $k$-th pivot point in $\upseq$.

\noindent\textbf{Providers}.
The pivot abstraction simulates a run $\run$ under the TSO semantics by running abstract processes called \emph{providers} in a sequential manner.
For  $1 \leq k \leq \sizeof\upseq+1$, the $k$-provider simulates the process that generates the write of the rank $k$ message $\tuple\xd$ corresponding to the $k$-pivot in $\run$.
The $k$-provider completes its task when it has simulated this process until the point it generates $\tuple\xd$.
At this point, it invokes the $(k+1)$-provider.
With this background, we now develop the formal pivot semantics for parameterized $\tso\of{\adt}$.

\paragraph*{Formal Pivot semantics for Parameterized $\tso\of{\adt}$.}
We define the formal operational semantics of the pivot abstraction as a labelled transition system.
Given a process description $\process = \tuple{ \stateset, \state_\init, \transition }$ and ADT $\adt = \tuple{\vals, \set{\val_\init}, \ops, \step\adt}$,
a configuration of the pivot transition system represents the \emph{view} of a provider when simulating a run of the program.
A view $\view = \tuple{ \state, \val, \lastwrite, \upseq, \eptr, \lptr, \pptr }$ is defined as follows. %
The process state is given by $\state \in \stateset$.
The value of the provider's ADT $\adt$ is $\val \in \vals$.
The function $\lastwrite: \varset \to \valset\mathop\cup\set\none$ gives for each $\xvar\in \varset$, the value of the latest (i.e., most recent) write the provider has performed on $\xvar$.
If no such instruction exists (the process has made no writes to $\xvar$) then $\lastwrite(\xvar) = \none$.
Note that $\lastwrite$ abstracts the buffer in terms of read-own-write operations since the process can only read from the most recent pending write in its buffer on each variable (if it exists).
We define $\lastwrite_\none$ such that $\lastwrite_\none\of\xvar=\none$ for all $\xvar\in\varset$.
The first update sequence of pivot messages is $\upseq \in \msgs_\diff$.
It is unchanged by transitions and remains constant throughout the pivot run.

The \emph{external pointer}, $\eptr \in \set{0,1,\ldots,\sizeof\upseq}$ helps the provider keep track of which messages from $\upseq$ it has observed.
These messages have been  propagated by other processes.
The external pointer is used to  identify which variables are still holding their initial values in the memory.
If the provider observes an external write on a variable $x$ (by accessing the memory), then this write has overwritten the initial value of $x$ in the memory.
The \emph{local pointer} $\lptr: \varset \to \set{0,1,\ldots,\sizeof\upseq}$ is a set of pointers, one for each variable $\xvar\in\varset$.
The function $\lptr\of\xvar$ gives the highest ranked write operation the provider itself has performed (on any variable) before it performed the latest write on $\xvar$.
The local pointer is necessary to know which variables lose their initial values when we need to empty the buffer. In other words, the local pointer abstracts the buffer in terms of update operations.
We define $\maxlptr:=\max\set{\lptr\of\xvar \mid \xvar\in\varset}$ as the highest value of a local pointer and $\lptr^0$ such that $\lptr^0\of\xvar=0$ for all variables $\xvar\in\varset$, i.e., the pointers are all in the leftmost position.
The \emph{progress pointer} $\pptr \in \set{1,2,\ldots,\sizeof\upseq+1}$ gives the rank of the process the current provider is simulating.

\begin{figure}[ht]
\openup 4pt
\begin{gather*}
    \scaledInference[skip]{
        \tuple{ \state, \skip, \state' } \in \transition
    }{
        \tuple{ \state, \val, \lastwrite, \upseq, \eptr, \lptr, \pptr } \step[\skip]\pvt \tuple{ \state', \val, \lastwrite, \upseq, \eptr, \lptr, \pptr }
    }
    \\
    \scaledInference[write(1)]{
        \tuple{ \state, \wr\of\xd, \state' } \in \transition, \pos\of\upseq\of{\tuple\xd} < \pptr, \lptr' = \lptr[\xvar \leftarrow \max\of{\pos\of\upseq\of{\tuple\xd}, \maxlptr}]
    }{
        \tuple{ \state, \val, \lastwrite, \upseq, \eptr, \lptr, \pptr } \step[\wr\of\xd]\pvt \tuple{ \state, \val, \lastwrite[\xvar \leftarrow \dval], \upseq, \eptr, \lptr', \pptr }
    }
    \\
    \scaledInference[write(2)]{
        \tuple{ \state, \wr\xd, \state' } \in \transition, \pos\of\upseq\of{\tuple\xd} = \pptr
    }{
        \tuple{ \state, \val, \lastwrite, \upseq, \eptr, \lptr, \pptr } \step[\wr\of\xd]\pvt \view_\init\of{\upseq, \pptr+1}
    }
    \\
    \scaledInference[read(1)]{
        \tuple{ \state, \rd\of\xd, \state' } \in \transition, \lastwrite\of\xvar = \dval
    }{
        \tuple{ \state, \val, \lastwrite, \upseq, \eptr, \lptr, \pptr } \step[\rd\of\xd]\pvt \tuple{ \state', \val, \lastwrite, \upseq, \eptr, \lptr, \pptr }
    }
    \\
    \scaledInference[read(2)]{
        \tuple{ \state, \rd\of\xd, \state' } \in \transition, \dval = \init\of\xvar, \lastwrite\of\xvar = \bot, \pos\of\upseq\of\xvar > \eptr
    }{
        \tuple{ \state, \val, \lastwrite, \upseq, \eptr, \lptr, \pptr } \step[\rd\of\xd]\pvt \tuple{ \state', \val, \lastwrite, \upseq, \eptr, \lptr, \pptr }
    }
    \\
    \scaledInference[read(3)]{
        \tuple{ \state, \rd\of\xd, \state' } \in \transition, \pos\of\upseq\of{\tuple\xd} < \pptr, \eptr' = \max\of{\eptr, \lptr\of\xvar, \pos\of\upseq\of{\tuple\xd}}
    }{
        \tuple{ \state, \val, \lastwrite, \upseq, \eptr, \lptr, \pptr } \step[\rd\of\xd]\pvt \tuple{ \state', \val, \lastwrite, \upseq, \eptr', \lptr, \pptr }
    }
    \\
    \scaledInference[memory-fence]{
        \tuple{ \state, \mf, \state' } \in \transition, \eptr' = \max\of{\eptr, \maxlptr}
    }{
        \tuple{ \state, \val, \lastwrite, \upseq, \eptr, \lptr, \pptr } \step[\mf]\pvt \tuple{ \state', \val, \lastwrite, \upseq, \eptr', \lptr, \pptr }
    }
    \\
    \scaledInference[data-operation]{
        \tuple{ \state, \op, \state' } \in \transition, \op \in \ops, \val \step[\op]\adt \val'
    }{
        \tuple{ \state, \val, \lastwrite, \upseq, \eptr, \lptr, \pptr } \step[\op]\pvt \tuple{ \state', \val', \lastwrite, \upseq, \eptr, \lptr, \pptr }
    }
\end{gather*}
\caption{The transition relation of the pivot semantics for a process $\process$.}
\label{fig:rules_pivot}
\end{figure}

Given an update sequence $\upseq \in \msgs^\diff$ and $1 \leq k \leq \sizeof\upseq+1$, we define \emph{the initial view induced by $\upseq$ and $k$} denoted by $\view_\init(\upseq, k)$, as the view $\tuple{ \state^\init, \val_\init, \lastwrite_\bot, \upseq, 0, \lptr^0 , k }$.
For a given $\upseq$, the $k$-provider starts with $\view_\init(\upseq, k)$: $\lastwrite_\bot$ and $\lptr^0$ imply that the simulated process has not performed any writes and $\eptr = 0$ means that it has not read/updated from/to the memory.

We define the labeled transition relation $\step\pvt$ on the set of views by the inference rules given in \autoref{fig:rules_pivot}.
The set of labels is $\instrs \mathop\cup \ops$.
We describe the inference rules briefly. %
The $\skip$ rule only changes the local state of the process.
There are two inference rules, write(1) and write(2), to describe the execution of a write operation $\wr\of\xd$.
The rule write(1) describes the situation when the rank of $\tuple\xd$ is strictly smaller than the progress pointer $\pptr$.
In this case, we update both $\lastwrite$ and $\lptr$.
The rule write(2) describes the situation when the rank of $\tuple\xd$ equals the progress pointer.
This means that the provider has provided the message $\tuple\xd$ with rank $\pptr$.
Hence it has completed its mission, and initiates the next provider by transitioning to $\view_\init(\upseq, \pptr + 1)$.

There are three inference rules that describe a read operation $\rd\of\xd$.
The rule read(1) describes when the last written value to $\xvar$ by the provider is $\dval$, $\lastwrite\of\xvar = \dval$.
In this case, the provider simply reads from its local buffer.
The rule read(2) describes the read of an initial value.
It ensures that the read is possible by checking that no write operation on $\xvar$ is executed by the provider ($\lastwrite\of\xvar = \bot$), and by checking that the initial value of the variable has not been overwritten in the memory.
This is achieved by checking if the position of $\tuple\xd$ in $\upseq$, i.e. $\pos\of\upseq(\tuple\xd)$, is strictly larger than $\eptr$.
The rule read(3) describes when the simulated process reads from the memory.
It checks that the message $\tuple\xd$ has been generated by some previous provider ($\pos\of\upseq(\tuple\xd) < \pptr$), and then it updates the external pointer to $\max(\eptr, \lptr\of\xvar, \pos\of\upseq(\tuple\xd))$.
The memory fence rule describes when the simulated process does a fence action.
The rule updates the external pointer to $\max(\eptr, \maxlptr)$.
Finally, the data-operation rule describes when the simulated process does an ADT operation.

The set of initial views is $\viewset_\init = \set{ \view_\init(\upseq, 1) \mid \upseq \in \msgs^\diff }$.
This is the set of initial views of the $1$-provider and it is finite because $\msgs^\diff$ is finite, unlike the set of initial configurations $\confset_\init$ in the parameterized case under TSO.
\section{Register Machines}
Our goal is to design a general method to determine the decidability and complexity of \tsopreach\ depending on $\adt$.
We examine the pivot abstraction introduced in the previous chapter.
A view $\view = \tuple{ \state, \val, \lastwrite, \upseq, \eptr, \lptr, \pptr }$
of the pivot transition system, can be partitioned into the following two components:
(1) $\state, \lastwrite, \upseq, \eptr, \lptr, \pptr$
which contains the local state and also effectively abstracts the unbounded FIFO buffers and shared memory of the TSO system and
(2) $\val$ which captures the value of the ADT.
The first part is finite since each component takes finitely
many values. We call this the \emph{book-keeping} state since it keeps track of the
progress of the core TSO system.
However, the ADT part can be infinite, depending upon
the abstract data type.

We will use a register machine in order to represent the book-keeping state
in a finite way using states and registers.
On the other hand, we will keep the ADT component general
and only later instantiate it to some interesting cases.

A \textit{register machine} is a finite state automaton that has access to a finite set of \textit{registers}, each holding a natural number.
The register machine can execute two operations on a register, it can write a given value or it can read a given value.
A read is blocking if the given value is not in the register.
We differ from most definitions of register machines in two significant ways:
Since we only require a finite domain to model $\tso\of{\adt}$ semantics, the values of the registers are bound from above by an $N \in \mathbb{N}$.
This makes the register assignments finite whereas most definitions allow for an unbounded domain.
Further, our register machine is augmented with an ADT.

Given an ADT $\adt = \tuple{ \vals, \set{\val_\init}, \ops, \step\adt}$, let $\regs$ be a finite set of registers and $\valset = \{0, \dots, N\}$ their domain.
We define the set of actions $\acts = \set{ \SKP, \WRITE\of\rgd, \READ\of\rgd \mid \rg \in \regs, \dval \in \valset }$.
A register machine is then defined as a tuple $\RM = \tuple{ \stateset, \state_\init, \transition }$, where $\stateset$ is a finite set of states, $\state_\init \in \stateset$ is the initial state and $\transition \subseteq \stateset \times (\acts \cup \ops) \times \stateset$ is the transition relation.

The semantics of the register machine are given in terms of a transition system.
The set of configurations is $\stateset \times \valset^\regs \times \vals$.
A configuration consists of a state, a register assignment $\regs \to \valset$ and a value of $\adt$.
The initial configuration is $\tuple{\state_\init, 0^\regs, \val_\init}$, where all registers contain the value $0$.

The transition relation $\to$ is described in the following.
$\SKP$ only changes the local state, not the registers or the ADT value.
$\WRITE\of\rgd$ sets the value of the register $\rg$ to $\dval$.
$\READ\of\rgd$ is only enabled if the value of $\rg$ is $\dval$, it does not change the value.
The operations in $\ops$ work as usually, they do not change any register.
We define the state reachability problem for register machines as \rmreach\ in the usual way.
A state $\state_\final\in\stateset$ is reachable if there is a run of the transition system defined by the semantics of $\RM$ that starts in the initial configuration and ends in a configuration with state $\state_\final$.

\subsection{Simulating Pivot Abstraction by Register Machines}\label{sec:simulatepivotreg}
In this section we will show how to simulate the pivot abstraction by a register machine. The idea is to save the book-keeping state (except for the local state) in the registers. Given a process description $\process = \tuple{ \stateset^\process, \state_\init^\process, \transition^\process }$ for an ADT $\adt$, we construct a register machine $\RM = \tuple{\stateset, \state_\init,\transition }$ that simulates the pivot semantics as follows. The set of registers is
$$\resizebox{1\hsize}{!}{$ \regs := \set{ \lastwrite\of\xvar, \rank_\varset\of\xvar, \rank_\msgs\of\msg, \eptr, \lptr\of\xvar, \maxlptr, \pptr, \rank_\mathsf{nxt} \mid \xvar \in \varset, \msg \in \msgs }~.$ }$$
The registers $\rank_\varset\of\xvar$ and $\rank_\msgs\of\msg$ hold the rank of each variable and message, respectively. This implicitly gives rise to an update sequence.
The auxiliary register $\rank_\mathsf{nxt}$ is used to initialize the other rank registers, as will be explained later on. The remaining registers correspond to their respective counterparts in the pivot abstraction. Note that the number of registers is linear in the number of messages $\sizeof\msgs$.
The domain of the registers is defined to be $\valset = \set{ 0, \dots, \sizeof\msgs + 1 }$. Since the TSO memory domain is finite, we can assume w.l.o.g. that the memory values are positive integers. If $\lastwrite\of\xvar = 0$, it means that there has been no write on $\xvar$ and it still holds the initial value. The set of states $\stateset$ contains  $\stateset^\process \cup \{\state_\init^\RM, \state_\init^\mathsf{ptr}\}$ as well as a number of (unnamed) auxiliary states that will be used in the following.

To simplify our construction, we will use additional operations on registers, instead of just $\WRITE$ and $\READ$. We introduce different blocking comparisons between registers and values such as $==,<,\leq,\neq$, register assignments such as $\rg:=\rg'$, and increments by one denoted as $\rg\texttt{++}$.
A more detailed description of these instructions is given in Appendix~\ref{app:register}.

\textbf{The Initializer.}
The pivot semantics define an exponential number of initial states: one per possible update sequence.
The register machine instead guesses an update sequence at the start of the execution and stores it in the rank registers. This part of the register machine is the \textit{rank initializer} (shown in \autoref{fig:rm_simulator} (a)). It uses the auxiliary register $\rank_\mathsf{nxt}$ to keep track of the next rank that is to be assigned.  In a nondeterministic  manner, the rank initializer chooses a so far unranked message and then it assigns the next rank to this message.
If the variable of the message has no rank assigned yet, it updates the rank of the variable.
Then it increases the $\rank_\mathsf{nxt}$ register and continues. After each rank assignment, the initializer can choose to stop the rank assignment. In that case, it initializes the register $\pptr$ to 1 and finishes in the initial state of $\process$.

In addition to the rank initializer, we have the \textit{pointer initializer}. 
It is responsible for resetting all pointers except the process pointer to zero. The process pointer is incremented by one instead. This initializer is not executed in the beginning of the simulation, but between epochs of the pivot abstraction.

%
%
%
%
%
%
%

\textbf{The simulator.} The main part of this construction handles the simulation of the pivot abstraction. It contains $\stateset^\process$ as well as several auxiliary states that are described in the following. It simulates each instruction of $\tso\of\adt$.
The skip instruction and the data instructions are carried out unchanged.
A visualization of the remaining instructions is depicted in \autoref{fig:rm_simulator}.
In case of a write instruction $\wr\of\xd$, we first compare the rank of the write message with the process pointer. If they are equal, it means that the epoch is finished and the next process should start, therefore we jump to the first state of the pointer initializer. Otherwise, we set the last write pointer $\lastwrite\of\xvar$ to $\dval$. Now, we ensure that $\lptr^\mathsf{max}$ is at least as large as the rank of $\tuple\xd$ and finally we update the local pointer $\lptr\of\xvar$ to be equal to $\lptr^\mathsf{max}$.
For the memory fence instruction, it only needs to be ensured that the external pointer is at least as large as the maximum local pointer $\lptr^\mathsf{max}$.
For a read instruction $\rd\of\xd$, if the last write to $\xvar$ was of value $\dval$, we can execute the read directly. Otherwise, after checking that the write can be performed by the current provider, we ensure that the external pointer is at least as large as both the rank of $\tuple\xd$ and the local pointer of $\xvar$. For the special case that $\dval = \dval_\init$, there is an additional way in which the read can be performed: We can read $\dval_\init$ from the memory if the process has neither already written to $\xvar$ nor observed a write that has higher or equal rank than the rank of $\xvar$.
This gives us the following theorem, proven in Appendix~\ref{app:register}:
\begin{theorem}\label{thm:cmplx_pivot_register}
	\tsopreach\ is polynomial time reducible to \rmreach.
\end{theorem}
\begin{figure}[h]
	\centering
	\resizebox{\textwidth}{!}{
		\subfloat[width=\textwidth][The rank initializer]{
		\begin{tikzpicture}[
			state/.style={circle, draw=black, minimum size=16pt},
			scale=2.5,yscale=-0.7
			]
			\node[state]	at (1,0)	(q2) {};
			\node[state]	at (2,-1)	(q3) {};
			\node[state]	at (3,0)	(q4) {};

			\node[state]	at (3,-1)	(qh) {};

			\node			at (0,1) (r0) {$\state_\init^{\RM\of\process}$};
			\node[state]		at (2,1) (r1) {};
			\node			at (4,1) (r2) {$\state_\init^\process$};

			\draw[->]	(r1) -- node[below left]	{$\rank_\msgs\of{\tuple\xd} == 0$}		(q2);
			\draw[->]	(q2) -- node[above left,xshift=5]	{$\rank_\msgs\of{\tuple\xd} := \rank_\mathsf{nxt}$}	(q3);
			\draw[->]	(q3) -- node[below,rotate=-35]	{$\rank_\varset\of\xvar \neq 0$}					(q4);
			\draw[->]	(q4) -- node[above left]	{$\rank_\mathsf{nxt}\texttt{++}$}	(r1);

			\draw[->]	(q3) -- node[above]	{$\rank_\varset\of\xvar == 0$}					(qh);
			\draw[->]	(qh) -- node[right]	{$\rank_\varset\of\xvar := \rank_\mathsf{nxt}$}	(q4);

			\draw[->]	(r0) -- node[below] {$\rank_\mathsf{nxt} := 1$} (r1);
			\draw[->]	(r1) -- node[below] {$\pptr := 1$} (r2);

			\draw[thick,decoration={brace,mirror,raise=155pt},decorate] (r1.north) -- node[right=160pt] {$\forall\: \tuple\xd \in \msgs$} ($(q3.north)+(0,-0.25)$);

			\draw[ultra thick, dashed, black!50] ($(r1.north)+(-1.7,0)$) -- ($(r1.north)+(-1.7,-2.25)$) -- ($(r1.north)+(2,-2.25)$) -- ($(r1.north)+(2,0)$) -- cycle;
		\end{tikzpicture}
	}
}
\newline
		\resizebox{\hsize}{!}{
\subfloat[width=.9\textwidth][$\wr\of\xd$]{
	\begin{tikzpicture}[
		state/.style={circle, draw=black, minimum size=16pt},
		scale=3,yscale=-0.5
		]
		\node			at (0.2,0.4) (r0) {$\state$};
		\node[state]		at (1,0) (r1) {};
		\node[state]		at (2,0) (r2) {};
		\node[state]		at (3,0) (r3) {};
		\node			at (4,0) (r4) {$\state'$};

		\node			at (1,1) (s0) {$\state_\init^\mathsf{ptr}$};
		\node[state]		at (2.5,0.8) (s1) {};

		\draw[->]	(r0) -- node[above,rotate=14] {$\rank(\tuple\xd) < \pptr$} (r1);
		\draw[->]	(r1) -- node[above] {$\lastwrite\of\xvar := d$} (r2);
		\draw[->]	(r2) -- node[above] {$\lptr^\mathsf{max} \geq \rank(\tuple\xd)$} (r3);
		\draw[->]	(r3) -- node[above] {$\lptr\of\xvar := \lptr^\mathsf{max}$} (r4);

		\draw[->]	(r0) -- node[below,rotate=-21] {$\rank(\tuple\xd) == \pptr$} (s0);
		\draw[->]	(r2) -- node[left] {$\lptr^\mathsf{max} < \rank(\tuple\xd)$} (s1);
		\draw[->]	(s1) -- node[right] {$\lptr^\mathsf{max} := \rank(\tuple\xd)$} (r3);

	\end{tikzpicture}
}
\newline
\subfloat[width=.4\textwidth][$\mf$]{
	\begin{tikzpicture}[
		state/.style={circle, draw=black, minimum size=16pt},
		scale=3,
		]
		\node			at (0,0) (r0) {$\state$};
		\node[state]		at (0.5,0.5) (r1) {};
		\node			at (1,0) (r2) {$\state'$};

		\draw[->]	(r0) -- node[above,rotate=45] {$\eptr < \lptr^\mathsf{max}$} (r1);
		\draw[->]	(r1) -- node[above,rotate=-45] {$\eptr := \lptr^\mathsf{max}$} (r2);
		\draw[->]	(r0) -- node[below] {$\eptr \geq \lptr^\mathsf{max}$} (r2);
	\end{tikzpicture}
}
}		\resizebox{\hsize}{!}{
\subfloat[width=.9\textwidth][$\rd\of\xd$]{
\begin{tikzpicture}[
	state/.style={circle, draw=black, minimum size=16pt},
	scale=2,yscale=-0.9
]
	\node			at (1,0) (r0) {$\state$};
	\node[state]		at (2,0) (r1) {};
	\node[state]		at (4,0) (r2) {};
	\node			at (6,0) (r3) {$\state'$};

	\node[state]		at (3,1) (s0) {};
	\node[state]		at (5,1) (s1) {};

	\draw[->]	(r0) -- node[below,yshift=-5] {$\rank(\tuple\xd) < \pptr$} (r1);
	\draw[->]	(r1) -- node[above] {$\eptr \geq \lptr\of\xvar$} (r2);
	\draw[->]	(r2) -- node[above] {$\eptr \geq \rank(\tuple\xd)$} (r3);

	\draw[->]	(r1) -- node[below,rotate=-45] {$\eptr < \lptr\of\xvar$} (s0);
	\draw[->]	(s0) -- node[above,rotate=45] {$\eptr := \lptr\of\xvar$} (r2);
	\draw[->]	(r2) -- node[below,rotate=-45] {$\eptr < \rank(\tuple\xd)$} (s1);
	\draw[->]	(s1) -- node[below,rotate=45] {$\eptr < \rank(\tuple\xd)$} (r3);

	\draw[->]	(r0) to[bend right] node[below] {$\lastwrite\of\xvar == d$} (r3);
\end{tikzpicture}
}
\subfloat[width=.4\textwidth][$\rd(\xvar, \dval_\init)$]{
\begin{tikzpicture}[
	state/.style={circle, draw=black, minimum size=16pt},
	scale=3
]
	\node			at (0,0) (r0) {$\state$};
	\node[state]		at (0.5,0.5) (r1) {};
	\node			at (1,0) (r2) {$\state'$};

	\draw[->]	(r0) -- node[above,rotate=45] {$\lastwrite\of\xvar == 0$} (r1);
	\draw[->]	(r1) -- node[above,rotate=-45] {$\rank\of\xvar > \eptr$} (r2);
\end{tikzpicture}
}}
\caption{The rank initializer and the simulator for some instructions $\instr$.}
\label{fig:rm_simulator}
\vspace{-5mm}
\end{figure}

\subsection{Simulating Register Machines by TSO}
We will now show how to simulate an ADT register machine with a parameterized program running under $\tso\of{\adt}$. The main idea is to save the information about the  registers in the last pending write operations, while making sure that not a single write operation actually hits the memory. Thus, the simulator always reads the initial value or its own writes, never writes of other processes.

The TSO program has a variable for each register, and two additional variables $\xvar_s$ and $\xvar_c$ that act as flags: $\xvar_s$ indicates that the verifier should start working, while $\xvar_c$ indicates that the verifier has successfully completed the verification.
At the beginning of the execution, each process nondeterministically chooses to be either \emph{simulator}, \emph{scheduler}, or \emph{verifier}. Each role will be described in the following. The complete construction is shown in Appendix~\ref{sec:app_regtotso}.

The simulator uses the same states and transitions as $\RM$, but instead of reading from and writing to registers, it uses the memory. If the simulator reaches the target state $\state_\target$, it first checks the $\xvar_s$ flag. If it is already set, the simulator stops, never reaching the final state $\state_\final$.
Otherwise, it waits until it observes the flag $\xvar_c$ to be set. It then enters the final state.
The scheduler's only responsibility is to signal the start of the verification process. It does so by setting the flag $\xvar_s$ at a nondeterministically chosen time during the execution of the program.
The verifier waits until it observers the flag $\xvar_s$. It then starts the verification process, which consists of checking each variable that corresponds to a register. If all of them still contain their initial value, the verification was successful. The verifier signals this to the simulator process by setting the $\xvar_c$ flag.

Any execution ending in $\state_\final$ must perform a simulation of $\RM$ ending in $\state_\target$ first, then a scheduler propagates the setting of flag $\xvar_s$ and afterwards a verifier executes. This ensures that the initial values are read by the verifier after the register machine has been simulated and thus the shared memory is unchanged. This means the simulator only accessed its write buffer and not writes from other threads.
It follows that $\state_\target$ is reachable by $\RM$ if and only if $\state_\final$ is reachable by $\process$ under $\tso\of{\adt}$.
This gives us the following result:

\begin{theorem}\label{thm:hard_pivot_register}
\rmreach\ is polynomial time reducible to \tsopreach.
\end{theorem}

\autoref{thm:cmplx_pivot_register} and \autoref{thm:hard_pivot_register} give us a method of determining upper and lower bounds of the complexity of $\tsopreach$ for different instantiations of ADT.
Since we have reductions in both directions, we can conclude that \\$\tsopreach$ is decidable if and only if $\rmreach$ is decidable.
	We know \tsopreach\ is \pspace-hard for \tsopreachof\noadt\ where $\noadt$ is the trivial ADT that models plain TSO semantics \cite{par_atig}.
	We can immediately derive a lower bound for any ADT:
	\tsopreach\  is \pspace-hard.

\section{Instantiations of ADTs}
In the following, we instantiate our framework to a number of ADTs in order to show its applicability.

\begin{theorem}\label{thm:1counter}
		\tsopreachof{\counter} and \tsopreachof{\wkcounter} are \pspace-complete.
\end{theorem}

We know \tsopreach\ is \pspace-hard for any ADT $\adt$ including \counter\ and \wkcounter.
Regarding the upper bound for $\counter$, we can show that \rmreachof{\counter} can be polynomially reduced to \rmreachof{\noadt}.
The idea is to show that there is a bound on the counter values in order to find a witness for  \rmreachof{\counter}. This bound is polynomial in the number of possible states and register assignments (i.e., this bound is at most exponential in the size of $\RMof{\counter}$.)
Assume a run that contains a configuration $\conf$ with a value that exceeds the bound, then certain state and register assignment are repeated in the run with different values. We can use this to shorten the run such that the counter value in $\conf$ is reduced.

We can encode the counter value (up to this bound) in a binary way into registers acting as bits. The  number of additional registers is polynomial in the size of $\RMof{\counter}$.
In order to simulate an $\inc$ operation on this binary encoding using $\WRITE$ and $\READ$, we only have to go through the bits starting at the least important bit and flip them until one is flipped from $0$ to $1$.
The $\dec$ operation works analogously. This only requires a polynomial state and transition overhead.

We know that \rmreachof{\noadt} is in \pspace \cite{par_atig}.
It follows from the polynomial reduction that \rmreachof{\counter} is in \pspace.
Applying \autoref{thm:cmplx_pivot_register} gives us that \tsopreachof{\counter} is in \pspace.
Since any \wkcounter\ is  a \counter, it follows  \tsopreachof{\wkcounter} is in \pspace\ as well.  The  proof can be found in Appendix~\ref{app:counter}.

\begin{theorem}\label{thm:reach_stack}
    \tsopreachof\stack\ is \exptime-complete.
\end{theorem}
    For membership, we encode the registers of $\RMof\stack$ in the states, which yields a finite state machine with access to a stack, i.e. a pushdown automaton. The construction has an exponential number of states. From \cite{Abo}, we have that checking the emptiness of a context-free language generated by a pushdown automaton is polynomial in terms of the size of the automaton. Combined, we get that state reachability of the constructed pushdown automaton is in \exptime.
    It follows that \rmreachof{\stack} is in \exptime\ (thanks to \autoref{thm:cmplx_pivot_register}).

To prove the lower bound, we can reduce the problem of checking the emptiness
of the intersection of a pushdown with $n$ finite-state automata \cite{Hue} to  \rmreachof{\stack}. This problem is well-known to be  \exptime-complete. The idea is to use the stack to simulate pushdown automaton and  $n$ registers to keep track of the  states of  the finite-state automata.
We apply \autoref{thm:hard_pivot_register} and get \tsopreachof{\stack} is \exptime-hard.
The  full proof  can be found in Appendix~\ref{app:stack}.

\begin{theorem}\label{thm:petri}
	\tsopreachof{\petri} is \expspace-complete.
\end{theorem}
\begin{proof}
Petri net coverability is known to be \expspace\ complete \cite{petri}.
We show hardness by reducing coverability of a marking $m$ to \rmreachof{\petri}. The idea is to  construct a register machine with a Petri net as ADT. This register machine will have two states $\state_\init $ and $\state_\final$. For every transition $t$ of the original Petri net, we have $t$: $\state_\init \step[t]{} \state_\init$ as a transition of the register machine (we simply simulate the original Petri net). Furthermore, we have  $\state_\init \step[t_{-m}]{} \state_\final$
as a transition of the register machine.
Thus, the state $\state_\final$ can be reached iff $m$ can be covered.

We reduce reachability of $\RMof{\petri}$ to Petri net coverability.
We construct the Petri net by taking the ADT $\petri$ and adding a place $p_\state$ for every state $\state$ and a place $p_{\reg,d}$ for every register $\reg\in\regs$ and register value $d\in \valset$.
The idea is that a marking with a token in $p_\state$ and one in $p_{\reg,d}$ but none $p_{\reg,d'}$ for $d' \neq d$ corresponds to a configuration of $\RMof{\petri}$ with state $q$ and $\reg = d$.
The value of $\petri$ is given by the remainder of the marking.

We simulate any $\state \step[\instr]{} \state'$ with a transition $t$ that takes one token from $\state$ and puts one in $\state'$.
If $\instr\in \ops$, then $\instr$ is a Petri net transition. We simply add the same input and output arcs to $t$.
To simulate a write, we add a new transition $t_{d'}$ for every $d'\in\valset$ with an arc to $p_{\reg,d}$ and an arc from $p_{\reg,d'}$.
The initial marking is consistent with $\val_\init^\petri$ and has one token in $p_{\state_\init}$.
A state $\state$ is reachable if a marking with one token in $p_{\state}$ is coverable.
\end{proof}

\paragraph{Higher Order ADTs.}
Let \fsmreach\ problem be the restriction of  $\rmreach$ with no registers.
The \fsmreach\ problem has been studied for many ADT such as higher order counter and higher order stack variations\cite{10.1007/978-3-642-40313-2_47,ENGELFRIET199121}.

\begin{theorem}\label{thm:HOcomplexity}
	\hfil
	\begin{itemize}
		\item \tsopreachof{\nof{\stack}}\ is $\hoadt{(n-1)}{\exptime}$-hard and in $\nof{\exptime}$.
		\item \tsopreachof{\nof{\wkcounter}}\ is $\hoadt{(n-2)}{\exptime}$-hard and in $\hoadt{(n-1)}{\exptime}$. 
		\item \tsopreachof{\nof{\counter}}\ is $\hoadt{(n-2)}{\expspace}$-hard and in $\hoadt{(n-1)}{\expspace}$. 
	\end{itemize}
\end{theorem}
\begin{proof}
\fsmreachof{\nof{\stack}}\ has been shown to be $\hoadt{(n-1)}{\exptime}$-complete \cite{ENGELFRIET199121}. We know \fsmreachof{\nof{\wkcounter}}\ is $\hoadt{(n-2)}{\exptime}$-complete and \fsmreachof{\nof{\counter}}\ is  $\hoadt{(n-2)}{\expspace}$-complete \cite{10.1007/978-3-642-40313-2_47}.
Since the reduction from \fsmreach\ to \rmreach\ is trivial, any hardness result can be applied to \tsopreach\ immediately using \autoref{thm:hard_pivot_register}. In order to reduce \rmreach\ to \fsmreach, we  encode register assignments into the state which results in an exponential state explosion. Then we apply \autoref{thm:cmplx_pivot_register} to obtain our upper bound.
\end{proof}

\begin{theorem}
	\tsopreachof{\nof{\multistack}} is $\hoadt{2}{\etime}$-complete.
\end{theorem}
\begin{proof}
	We know that \fsmreachof{\nof{\multistack}} is $\hoadt{2}{\etime}$-complete \cite{10.1007/978-3-540-85780-8_9} and we can apply \autoref{thm:hard_pivot_register} to get $\hoadt{2}{\etime}$-hardness.
	According to Theroem 4.6 in \cite{lmcs:871}, \fsmreachof{\nof{\multistack}} is in $\mathcal{O}\of{\sizeof{\automaton\of\adt}^{2^{dn}}}$ for some constant $d\in \mathbb{N}$. We apply the exponential size reduction to \rmreachof{\nof{\multistack}} and \autoref{thm:cmplx_pivot_register} and get \tsopreachof{\nof{\multistack}} is in $\mathcal{O}\of{({2^{ \sizeof{\program} }})^{2^{dn}} } =\mathcal{O}\of{2^{\sizeof{\program} \cdot 2^{dn}} }$ and thus it is also in  $\mathcal{O}\of{2^{2^{\sizeof{\program}} \cdot 2^{dn}} }= \mathcal{O}\of{2^{2^{{\sizeof{\program}} +dn}} }$. Thus,  \tsopreachof{\nof{\multistack}} is in $\hoadt{2}{\etime}$.
\end{proof}

We study well structured ADTs \cite{FINKEL200163,Abdulla2000AlgorithmicAO} (defined in Appendix~\ref{app:wsts}):
\begin{theorem}
	If  ADT $\adt$ is well structured, then \tsopreach\ is decidable.
\end{theorem}
A register machine for a well structured ADT $\adt$ is equivalent to the composition of a well structured transition system (WSTS) modeling $\adt$ and a finite transition system (and thus a WSTS) that models  states and registers.
According to \cite{Abdulla2000AlgorithmicAO}, the composition is again a WSTS and reachability is decidable. The above theorem is then an immediate corollary of  \autoref{thm:cmplx_pivot_register}.

\section{Conclusions and Future Work}
In this paper, we have taken the first step to studying the complexity of
parameterized verification under weak memory models when the processes manipulate unbounded data domains.
Concretely, we have presented complexity results for
parameterized concurrent programs running on the classical
TSO memory model when the processes operate on an abstract data type.
We reduce the problem to reachability for register machines enriched with the given abstract data type.

State reachability for finite automata with ADT has been extensively studied for many ADTs\cite{10.1007/978-3-642-40313-2_47,ENGELFRIET199121}.
We have shown in \autoref{thm:HOcomplexity} that we can apply our framework to existing complexity results of this problem. 
This provides us with decidability and complexity results for the corresponding instances of \tsopreach.
However, due to the exponential number of register assignments, the upper bound is exponentially larger than the lower bound.
We aim to study these cases further and determine more refined parametric bounds.

A direction for future work is considering other memory models, such as the partial store ordering semantics, the release-acquire semantics, and the ARM semantics.
It is also interesting to re-consider the problem under the assumption of having
distinguished processes (so-called {\it leader processes}).
Adding leaders is known to make the parameterized verification problem harder.
The complexity/decidability of parameterized verification under TSO with a single leader is open, even when the processes are finite-state.
\bibliographystyle{plain}
\bibliography{bibdatabase}
\appendix
\newpage
\centerline{\large{Appendix}}
\section{Higher-order Stack}
\label{app:ho_stack}

We present higher order stacks \cite{DBLP:journals/corr/abs-0705-0262,ENGELFRIET199121}.
For a natural number $n > 0$ and finite set $A$, we define $A^n =(A^{n-1})^{*}$, where $A^0 = A$.
For $n > 1$, an $n$-order stack is a stack of $(n-1)$-order stacks, where a $1$-order stack is the standard stack as described in \autoref{ex:stack}. For a stack alphabet $\Gamma$, we define the $n$-order stack as follows. We define the set of values denoted by $\vals^n$ to be $\vals^n = \Gamma^n$ and we also assume that initially the $n$-order stack is empty and therefore we set initial value to be $\val_\init^n = \epsilon$.  The set of operations on the $n$-order stack is defined to be $\ops^n = \{\pop(\gamma), \push(\gamma), \pop_k, \push_k, \pop_n, \push_n ~  | ~ \gamma \in \Gamma, 2 \leq k < n\}$.  For a word $\omega = a_1 a_2 \hdots a_l$, where $\omega \in \Gamma^n$ and $a_i \in \Gamma^{n-1}$ for $1\leq i \leq l$, the transition relation on the $n$-order stack is denoted by $\adtstep{}{n}$ and it is defined as follows.
$$\omega \adtstep{\push(\gamma)}{n}  a_1 \hdots a_{l-1} a_l' ~ ~ \text{where} ~ ~ a_l  \adtstep{\push(\gamma)}{n-1} a_l'.$$
$$\omega \adtstep{\push_k}{n}  a_1 \hdots a_{l-1} a_l' ~ ~ \text{where} ~ ~ a_l  \adtstep{\push_k}{n-1} a_l'.$$
$$\omega \adtstep{\push_n}{n}  a_1 \hdots a_l a_l.$$
$$\omega \adtstep{\pop(\gamma)}{n}  a_1 \hdots a_{l-1} a_l' ~ ~ \text{where} ~ ~ a_l  \adtstep{\pop(\gamma)}{n-1} a_l'.$$
$$\omega \adtstep{\pop_k}{n}  a_1 \hdots a_{l-1} a_l' ~ ~ \text{where} ~ ~ a_l  \adtstep{\pop_k}{n-1} a_l'.$$
$$\omega \adtstep{\pop_n}{n}  a_1 \hdots a_{l-1}.$$
$$\omega \adtstep{\isEmpty}{n}  \omega ~ ~ \text{where} ~ ~ a_l \adtstep{\isEmpty}{n-1} a_l.$$
$$\omega \adtstep{\isEmpty_k}{n}  \omega ~ ~ \text{where} ~ ~ a_l \adtstep{\isEmpty_k}{n-1} a_l.$$
$$\omega \adtstep{\isEmpty_n}{n}  \omega ~ ~ \text{where} ~ ~ a_l = \epsilon.$$
Notice that the definition of the transition relation of the $n$-order stack depends on the definition of the lower order stacks. Furthermore, the base case which is the transition relation of the $1$-order stack is given in \autoref{ex:stack}.

\section{Semantics of $\tso\of\adt$}\label{sec:app:tso-semantics}
\noindent \textbf{Configurations}.
A TSO($\adt$) configuration is a tuple $\tuple{ \indexset, \statemap, \valuemap, \buffermap, \memorymap }$, where:
\begin{enumerate}
	\item $\statemap: \indexset \to \bigcup_{\pid \in \indexset} \stateset^\pid$ maps each process to its local state i.e $\statemap(\pid) \in \stateset^\pid$.
	\item $\valuemap: \indexset \to \vals$ represents the state of the ADT of each process.
	\item $\buffermap: \indexset \to \msgs\kstar$ represents the buffer state of each process.
	\item  $\memorymap: \varset \to \valset$ represents the memory state by storing the value of each shared variable.
\end{enumerate}
We define $\confset^\program$ to be the set of all possible configurations in TSO($\adt$) semantics.
The \emph{initial configuration} is $\conf_\init^\program = \tuple{ \indexset, \statemap_{\init}, \valuemap_{\init}, \buffermap_{\init}, \memorymap_{\init} }$, where $\statemap_\init\of\pid = \state^\pid_\init$, $\valuemap_\init\of\pid = \val_\init$, $\buffermap_\init\of\pid = \emptybuffer$, and $\memorymap_\init\of\xvar = \dval_\init$ for all $\pid \in \indexset$ and $\xvar \in \varset$.
That is, the set of initial configurations is the singleton $\confset_\init^\program = \set{ \conf_\init^\program }$, where each process is in its initial state, each ADT holds its initial value, each store buffer is empty and the memory holds the initial values (given by $\dval_\init$) of all variables.

\begin{figure}[h]
\openup 4pt
\begin{gather*}
    \inference[skip]{
        \pid \in \indexset, \tuple{ \statemap\of\pid, \skip, \state' } \in \transition^\pid
    }{
        \tuple{ \indexset, \statemap, \valuemap, \buffermap, \memorymap } \step[\tuple{ \pid, \skip }]\program \tuple{ \indexset, \statemap[\pid \leftarrow  \state'], \valuemap, \buffermap, \memorymap }
    }
    \\
    \inference[write]{
        \pid \in \indexset, \tuple{ \statemap\of\pid, \wr\of\xd, \state' } \in \transition^\pid
    }{
        \tuple{ \indexset, \statemap, \valuemap, \buffermap, \memorymap } \step[\tuple{ \pid, \wr\of\xd }]\program \tuple{ \indexset, \statemap[\pid \leftarrow \state'], \valuemap, \buffermap[\pid \leftarrow \tuple\xd \cdot \buffermap\of\pid], \memorymap }
    }
    \\
    \inference[read]{
        \pid \in \indexset, \tuple{ \statemap\of\pid, \rd\of\xd, \state' } \in \transition^\pid, \rval\of{\buffermap\of\pid, \memorymap\of\xvar} = \dval
    }{
        \tuple{ \indexset, \statemap, \valuemap, \buffermap, \memorymap } \step[\tuple{ \pid, \rd\of\xd }]\program \tuple{ \indexset, \statemap[\pid \leftarrow \state'], \valuemap, \buffermap, \memorymap }
    }
    \\
    \inference[memory-fence]{
        \pid \in \indexset, \tuple{ \statemap\of\pid, \mf, \state' } \in \transition^\pid, \buffermap\of\pid = \emptybuffer
    }{
        \tuple{ \indexset, \statemap, \valuemap, \buffermap, \memorymap } \step[\tuple{ \pid, \mf }]\program \tuple{ \indexset, \statemap[\pid \leftarrow \state'], \valuemap, \buffermap, \memorymap }
    }
    \\
    \inference[memory-update]{
        \pid \in \indexset, \buffermap\of\pid = \buffer \cdot \tuple\xd
    }{
        \tuple{ \indexset, \statemap, \valuemap, \buffermap, \memorymap } \step[\tuple{ \pid, \up\of\xd }]\program \tuple{ \indexset, \statemap, \valuemap, \buffermap[\pid \leftarrow \buffer], \memorymap[\xvar \leftarrow \dval] }
    }
    \\
    \inference[data-operation]{
        \pid \in \indexset, \op \in \ops, \tuple{ \statemap\of\pid, \op, \state' } \in \transition^\pid, \valuemap\of\pid \step[\op]\adt \val'
    }{
        \tuple{ \indexset, \statemap, \valuemap, \buffermap, \memorymap } \step[\tuple{ \pid, \op }]\program \tuple{ \indexset, \statemap[\pid \leftarrow \state'], \valuemap[\pid \leftarrow \val'], \buffermap, \memorymap }
    }
\end{gather*}
\caption{The transition relation of the TSO($\adt$) semantics.}
\label{fig:rules_tso}
\end{figure}

\noindent \textbf{Transition Relation and Labels}.
In \autoref{fig:rules_tso}, the transition relation of the TSO($\adt$) labelled transition system is given by a set of inference rules.
Each inference rule is annotated by a label $\lbl = \tuple{ \pid, \plbl }$, where $\pid \in \indexset$ is the index of the process executing the event and $\plbl$ is either a TSO instruction from $\instrs$, an update $\up\of\xd$ or an operation on the ADT denoted by $\op$.

Before we explain the inference rules, we need to introduce some auxiliary functions.
We define the (local value) function $\lval: \msgs\kstar \to \varset \to (\valset \cup \set\bot) $ such that $\lval\of\buffer\of\xvar = \dval$ if $\buffer = \word_1 \cdot \tuple\xd \cdot \word_2$ and $\word_1$ does not contain any assignment on the variable $\xvar$, or $\lval\of\buffer\of\xvar = \bot$ if $\buffer$ does not contain any assignment on $\xvar$.
Informally, given that $\buffer$ is the buffer state for some process, then $\lval\of\buffer\of\xvar$ returns the value of the most recent added pending message on $\xvar$ in $\buffer$, if such a message exists.
Otherwise it returns the special symbol $\bot$.
Note that we assume that messages are queued in from the left and taken out from the right.
Based on this, we define the (read value) function $\rval: (\msgs\kstar \times \valset) \to \varset \to \valset$ as $\rval\of{\beta,\dval}\of\xvar = \lval\of\beta\of\xvar$ if $\lval\of\beta\of\xvar \neq \bot$ and $\rval\of{\beta,\dval}\of\xvar = \dval$ otherwise.
Let $\dval$ be the value stored in the memory for variable $\xvar$.
The idea for $\rval$ is that the process will read its local value for $\xvar$ using $\lval$ if it exists. 
Otherwise, it will read the value $\dval$ directly from memory.
The TSO($\adt$) transition relation $\step\program$ is defined using the local transition relation $\transition_\pid$ of the individual processes $\process_\pid$.
Every rule, except for the memory-update, changes the local state of the process.
The write rule adds pending write message to the end of the buffer of the process.
The read rule uses the function $\rval$ to load the value of either the last pending message ($\lval\of\buffer\of\xvar \neq \bot$) on the considered variable, or, if the buffer contains no pending messages on the considered variable ($\lval\of\buffer\of\xvar = \bot$), it reads the value from the memory.
The memory-fence rule checks that the buffer is empty.
The memory-update rule can happen non-deterministically at any point in the run, and it removes the message at the head of the buffer and updates it to the memory without any effect on the local process state.
The data-operation rule describes the changes in local state of the ADT when a process performs an action of it.

A sequence of transitions $\conf_0 \step[\lbl_1]\program \conf_1 \step[\lbl_2]\program \conf_2 \dots \conf_{n-1} \step[\lbl_n]\program \conf_n$ where $\conf_0 = \conf_\init^\program$ is the initial configuration and $\conf_i$ is the configuration obtained from configuration $\conf_{i-1}$ using a rule $\lbl_i$ from \autoref{fig:rules_tso} is called a run in the TSO($\adt$) transition system. If there is a run ending in a configuration with state $\state_\final$, then we say $\state_\final$ is reachable by $\process$ under $\tso\of{\adt}$.
\section{Register Machines}\label{app:register}

Our definition of register machines uses only the instructions $\READ$ and $\WRITE$. We will call it instruction set I.
For convenience we introduce two more instruction sets. Table~\ref{tab:rm_actions} shows a list of all actions, grouped by instruction set, together with abbreviations that will be used throughout this section.

Set II contains the following instructions:
$\INC\of\rg$ and $\DEC\of\rg$ increases or decreases the value of the register $\rg$ by one, respectively.
The former is only enabled if the value of $\rg$ is less than the maximum value, the latter is only enabled if the value of $\rg$ is not equal to zero.
$\CKZ\of\rg$ does not change the register, but is only enabled if the value of $\rg$ equals zero.

It is easy to see that with these basic operations from set II, more complicated ones can be simulated. For example, it is possible to check for the equality of two registers, as shown in Figure~\ref{fig:rm_composite_action}. This construction uses one auxiliary register and has runtime overhead in the order of $|V|$.

\begin{figure}
	\centering
	\begin{tikzpicture}[
		state/.style={circle, draw=black, minimum size=16pt},
		scale=3,
		]
		\node        at (-1,0)	(s)  {};
		\node[state] at (0,0)	(l1) {};
		\node[state] at (0.5,0.8)	(d1) {};
		\node[state] at (-0.5,0.8)	(d2) {};
		\node[state] at (1,0)	(eq) {};
		\node[state] at (2,0)	(l2) {};
		\node[state] at (2.5,0.8)	(i1) {};
		\node[state] at (1.5,0.8)	(i2) {};
		\node        at (3,0)	(e)  {};

		\draw[->]	(s)  -- node[below]			{skip}			(l1);
		\draw[->]	(l1) -- node[below right]	{$\DEC(\rg_1)$}	(d1);
		\draw[->]	(d1) -- node[above]			{$\DEC(\rg_2)$}	(d2);
		\draw[->]	(d2) -- node[below left]		{$\INC(\rg_3)$}	(l1);
		\draw[->]	(l1) -- node[below]			{$\CKZ(\rg_1)$}	(eq);
		\draw[->]	(eq) -- node[below]			{$\CKZ(\rg_2)$}	(l2);
		\draw[->]	(l2) -- node[below right]	{$\DEC(\rg_3)$}		(i1);
		\draw[->]	(i1) -- node[above]			{$\INC(\rg_1)$}	(i2);
		\draw[->]	(i2) -- node[below left]		{$\INC(\rg_2)$}	(l2);
		\draw[->]	(l2) -- node[below]			{$\CKZ(\rg_3)$}	(e);
	\end{tikzpicture}
	\caption{Checking for equality of two registers $\rg_1$ and $\rg_2$, using only the operations $\INC$, $\DEC$ and $\CKZ$ and the auxiliary register $\rg_3$.}
	\label{fig:rm_composite_action}
\end{figure}

\begin{table}
	\centering
	\begin{tabular}{>{\centering}m{24pt}>{\centering}m{24pt}@{\quad}l@{\qquad}l@{\qquad}l}
		\hline
		\multicolumn{2}{c}{\textbf{model}} & \textbf{action} & \textbf{abbrv.} & \textbf{semantics} \\
		\hline
		\multirow{2}{*}{I}
		& & $\WRITE\of\rgd$ & $\rg := \dval$ & set value of $\rg$ to $\dval$ \\
		& & $\READ\of\rgd$ & $\rg == \dval$ & check if $\rg$ contains $\dval$ \\
		\hline
		\multirow{10}{*}{III} &
		\multirow{3}{*}{II}
		& $\INC(\rg)$ & $\rg\texttt{++}$ & increase value of $\rg$ by one \\
		& & $\DEC(\rg)$ & $\rg\texttt{--}$ & decrease value of $\rg$ by one \\
		& & $\CKZ(\rg)$ & $\rg == 0$ & check if $\rg$ contains zero \\
		\cline{2-5}
		& & $\SET(\rg, y)$ & $\rg := y$ & set value of $\rg$ to $y$ \\
		& & $\CKE(x, y)$ & $x == y$ & check for equality of $x$ and $y$ \\
		& & $\CKNE(x, y)$ & $x \neq y$ & check for non-equality of $x$ and $y$ \\
		& & $\CKL(x, y)$ & $x < y$ & check if $x$ is less than $y$ \\
		& & $\CKG(x, y)$ & $x > y$ & check if $x$ is greater than $y$ \\
		& & $\CKLE(x, y)$ & $x \leq y$ & check if $x$ is less or equal than $y$ \\
		& & $\CKGE(x, y)$ & $x \geq y$ & check if $x$ is greater or equal than $y$ \\
		\hline
	\end{tabular}
	\vspace{8pt}
	\caption{Actions of a register machine, their abbreviations and their semantics. \\ $\rg$ is a register, $\dval$ is a value, $x$ and $y$ can be both a register or a value.}
	\label{tab:rm_actions}
\end{table}

\begin{lemma}\label{lem:register_set2to1}
	There is a reduction of state reachability of a register machine with instruction set III to state reachability with set II.
	The reduction is polynomial in the size of the register machine and the largest value $\dval_\mathsf{max}$ used in an instruction.
\end{lemma}
\begin{proof}
	We show the following:
	Any instruction from set III can be simulated with set II. 
	If $x$ and $y$ are registers, this requires a fixed number of states per instruction and one additional register.
	If one is a value $\dval$ greater than $1$, then it requires 
	$\dval_\mathsf{max}$ states once and one register per instruction but no more than $\dval_\mathsf{max}$ registers altogether.
	The overhead of transitions is linear in the overhead of states.

	We show how to simulate $\CKE(\rg_1, \rg_2)$ in \autoref{fig:rm_composite_action}.
	We can simulate  $\SET(\rg_1, \rg_2)$ using additional register $\rg$ by setting $\rg$ and $\rg_1$ to 0 with $\rg\texttt{--}$ loops followed by $\CKZ$.
	Then we move the value of $\rg_2$ to $\rg$ by decreasing $\rg_2$ and increasing $\rg$ in a loop until $\rg_2$ is 0.
	Then we move the value of $\rg$ to $\rg_1$ and $\rg_2$ by decrasing $\rg$ and increasing $\rg_1$ and $\rg_2$ until $\rg$ is 0.

	Note that we can easily check $\rg\neq 0$ by decrementing and then incrementing $\rg$ once.
	We can easily simulate any of of the checks that compare two register by decreasing both registers in a loop until one is zero and then testing the other one  similar to \autoref{fig:rm_composite_action}.

	We can simulate $\SET\of\rgd$ by setting $\rg$ to $0$ and then increasing it $\dval$ times. This requires $\dval$ many states.
	We can compare a register to a value $\dval$ by trying to decrease it $\dval$ times and checking if or when it reaches $0$. This requires $\dval$ many states as well.

	If the value is $0$ or $1$, then this only requires a fixed number of states.
	For checks, we either check for zero or whether we can decrease once or twice.
	We can set something to $0$ with a decreasing loop followed by a check for zero. If we set to $1$, then this is followed by an increase.
	So for $0$ or $1$, these operations require only a fixed number of states.

	Alternatively we can increase an auxiliary register $|V|$ many times at the beginning of the computation and each time we reach a value $\dval$ that is used in an instruction, we save it to an additional register $\rg_\dval$ and replace $\dval$ in the instruction with $\rg_\dval$.

	Note that in any of these constructions, the outgoing transitions of any new state are bounded by a fixed value.
\end{proof}

\begin{lemma}\label{lem:register_set2to3}
	There is a reduction of state reachability of a register machine with instruction set III to state reachability with set I.
	The reduction is polynomial in the size of the register machine and the largest value $\dval_\mathsf{max}$ used in an instruction.
\end{lemma}
\begin{proof}
	We can simulate a register machine with instruction set II using set I.
	We simulate $\state \to[\INC\of\rg] \state'$ by attempting to read all values and then writing the successor: $\state \to[\READ\rgd] \state_\dval \to[\WRITE(\rg,\dval+1)] \state'$ for all $\dval\in \valset$.
	This requires $\sizeof{V}$ many new states for any such instruction.
	We simulate $\DEC\of\rg$ analogously and $\CKZ\of\rg$ with $\READ(\rg, 0)$.

	We can then apply \autoref{thm:cmplx_pivot_register} to the register machine with instruction set I.
\end{proof}

%

\subsection{Proof of  \autoref{thm:cmplx_pivot_register}}
We show that \tsopreach\ is polynomially reducible to \rmreach.

We show that the above construction of $\RM$ allows for a polynomial time reduction of \tsopreach\ to \rmreach. First, we argue correctness:
The  pointer initializer (illustrated in \autoref{fig:rm_pointer_initializer}) correctly sets up a new provider.
\begin{figure}
	\centering
	\resizebox{\textwidth}{!}{
		\begin{tikzpicture}[
			state/.style={circle, draw=black, minimum size=16pt},
			scale=2.5,
			]
			\node			at (1,0) (r1) {$\state_\init^\mathsf{ptr}$};
			\node[state]		at (2,0) (r2) {};
			\node[state]		at (3,0) (r3) {};
			\node[state]		at (4,0) (r4) {};
			\node[state]		at (5,0) (r5) {};
			\node[state]		at (6,0) (r6) {};
			\node[state]		at (7,0) (r7) {};
			\node			at (8,0) (r8) {$\state_\init^\process$};

			\draw[->]	(r1) -- node[below] {$\eptr := 0$} (r2);

			\draw[->]	(r2) -- node[below] {$\lptr(\xvar_1) := 0$} (r3);
			\draw[->, dotted]	(r3) -- node[below] {\dots} (r4);
			\draw[thick,decoration={brace,mirror,raise=20pt},decorate] (r2.center) -- node[below=24pt] {$\forall\: \xvar \in \varset$} (r4.center);

			\draw[->]	(r4) -- node[below] {$\maxlptr := 0$} (r5);

			\draw[->]	(r5) -- node[below] {$\phi_\lastwrite\of\xvar := 0$} (r6);
			\draw[->, dotted]	(r6) -- node[below] {\dots\vphantom{0}} (r7);
			\draw[thick,decoration={brace,mirror,raise=20pt},decorate] (r5.center) -- node[below=24pt] {$\forall\: \xvar \in \varset$} (r7.center);

			\draw[->]	(r7) -- node[below] {$\pptr\texttt{++}$} (r8);
		\end{tikzpicture}
	}
	\caption{The pointer initializer.}
	\label{fig:rm_pointer_initializer}
\end{figure}

According to \autoref{fig:rm_simulator}, it is easy to see that the rank initializer guesses an update sequence and
all instructions are according to the PTSO($\adt$) semantics.
We will omit a more formal proof of correctness this since it is straight forward.

The number of states and transitions is $\mathcal{O}(\sizeof{\msgs})$ for the rank initializer, $\mathcal{O}(\sizeof{\varset})$ for the pointer initializer and $\mathcal{O}(\sizeof{\confset} + \sizeof{\transition^\process})$ for the simulator. It is linear in largest parameter $\sizeof{\transition^\process}$.
The construction uses different actions than $\READ$ and $\WRITE$.
We can simulate each such action with a number of states that is linear in the size of the construction and the size of $\valset$.
Since both the number of possible ranks of an update and the possible values of reads and writes is bounded by the number of transitions $\sizeof{\transition^\process}$, we have shown that we only require a domain of that size to accurately represent the pivot semantics.
It follows that the reduction is polynomial.

\subsection{Simulating a Register Machine by TSO}\label{sec:app_regtotso}
In this section we will show how to simulate a register machine with ADT by a parameterized program running under TSO($\adt$). The main idea is to save the information of the registers in the shared variables, while making sure that not a single write operation actually hits the memory. Thus, the process that simulates the register machine always reads the initial memory value or its own writes, but never writes of other processes.

Let $\RM = \tuple{\stateset, \state_\init, \regs, \valset, \transition }$ be a register machine with some ADT. 
Given a target state $\state_\target \in\stateset$, we construct a process $\process = (\confset^\process, \conf_\init^\process, \transition)$ that simulates the register machine.
The set of states $\stateset$ contains $\stateset^\RM \cup \{\state_\init^\process, \state_\final\}$ and some unnamed auxiliary states. The TSO program has one variable for each register, and two additional variables $\xvar_s$ and $\xvar_c$ that act as flags: $\xvar_s$ indicates that the verifier should start working, while $\xvar_c$ indicates that the verifier has successfully completed the verification. The domain of the variables is $\valset$.

At the beginning of the execution, each process non-deterministically chooses to be one of the \textit{simulator}, \textit{scheduler} or \textit{verifier}. Each role will be described in the following. The complete process is shown in Figure~\ref{fig:rm_simulate_rm}.

The simulator uses the same states and transitions as $\RM$, but instead of reading from and writing to registers, it uses the memory instead. If the simulator reaches the target state $\state_\target$, it first checks the status of the $\xvar_s$ flag. If it is already set, the simulator stops, never reaching the final state $\state_\final$. Otherwise, it waits until it observes the flag $\xvar_c$ to be set. It then enters the final state.

The scheduler's only responsibility is to signal the start of the verification process. It does so by setting the flag $\xvar_s$ at a nondeterministically chosen time during the execution of the program.

The verifier waits until it observers the flag $\xvar_s$. It then starts the verification process, which consists of checking the value of each variable that corresponds to a register. If all of them still contain their initial value, the verification was successful. The verifier signals this to the simulator process by setting the $\xvar_c$ flag.

\begin{figure}
	\centering
	\begin{tikzpicture}[
		state/.style={circle, draw=black, minimum size=16pt},
		scale=2.3,
		]
		\node			at (0,0) (qi) {$\state_\init^\process$};

		\node			at (1,1) (s1) {$\state_\init^\RM$};
		\node			at (2,1) (s2) {$\state_\target$};
		\node[state]		at (3,1) (s3) {};
		\node			at (4,1) (s4) {$\state_\final$};

		\node[state]		at (1,0) (h1) {};
		\node[state]		at (2,0) (h2) {};

		\node[state]		at (1,-1) (v1) {};
		\node[state]		at (2,-1) (v2) {};
		\node[state]		at (3,-1) (v3) {};
		\node[state]		at (4,-1) (v4) {};
		\node[state]		at (5,-1) (v5) {};

		\draw[->]	(qi) -- node[above left] {$\skip$} (s1);
		\draw[->, dotted] (s1) to[out=30, in=210] (s2);
		\draw[->]	(s2) -- node[above] {$\rd(\xvar_s,0)$} (s3);
		\draw[->]	(s3) -- node[above] {$\rd(\xvar_c,1)$} (s4);
		\draw[ultra thick, dashed, black!50] (0.75,0.5) rectangle ++(1.5,1);

		\draw[->]	(qi) -- node[above] {$\skip$} (h1);
		\draw[->]	(h1) -- node[above] {$\wr(\xvar_s,1)$} (h2);

		\draw[->]	(qi) -- node[above right] {$\skip$} (v1);
		\draw[->]	(v1) -- node[above] {$\rd(\xvar_s,1)$} (v2);
		\draw[->]	(v2) -- node[above] {$\rd(\xvar_{\rg_1},\bot)$} (v3);
		\draw[->,dotted]	(v3) -- node[above] {\dots} (v4);
		\draw[->]	(v4) -- node[above] {$\wr(\xvar_c,1)$} (v5);
		\draw[thick,decoration={brace,mirror,raise=12pt},decorate] (v2.center) -- node[below=16pt] {$\forall\: \rg \in \regs$} (v4.center);
	\end{tikzpicture}
	\caption{Simulation of a register machine by a PTSO program.}
	\label{fig:rm_simulate_rm}
\end{figure}

\begin{theorem}
	If \rmreach\ is $\class$-hard, then so is \tsopreach.
\end{theorem}
\begin{proof}
	We show that there is a polynomial time reduction of \rmreach\ to \tsopreach.
	First, we show that the target state $\state_\target$ of $\RM$ is reachable if and only if the final state $\state_\final$ of $\process$ is reachable under PTSO.

	Any execution ending in $\state_\final$ must perform $\rd(\xvar_s,0)$ first, then the scheduler and afterwards the verifier. This ensures that the initial values are read by the verifier after the register machine has been simulated and thus the shared memory is unchanged. This means the simulator only accessed its write buffer and not writes from other threads.
	It is easy to see that we can always construct such an execution if $\state_\target$ is reachable by $\RM$.

	The construction requires $\sizeof{R}$ more states and thus its size is polynomial in the size of the register machine.
\end{proof}

\section{Reachability in TSO($\counter$)}
\label{app:counter}
\begin{proof}[of \autoref{thm:1counter}]
	We prove that \tsopreachof{\counter} is in \pspace.
	We show that \rmreachof{\counter} can be polynomially reduced to \rmreachof{\noadt}.
	We add a transition $\state_\final \step[\dec] {}\state_\final$.
	If a run reaches reaches $\state_\final$, we can extend this run with this new transition until the counter is $0$.
	This ensures that $\state_\final$ is reachable if and only if a configuration $\conf_\final$ is reachable that has state $\state_\final$ and counter value $\val_\final=0$.
	Note that this leaves reachability of $\state_\final$ unchanged.

	Let $n=\sizeof{Q\times \valset^\regs}$ be the number of possible states and register assignments.
	We show that any shortest run that reaches a configuration $\conf_{\final}$ with state $\state_\final$ and counter value $\val_\final=0$ only has counter values up to $n^2$.
	Assume there is a shortest run $\run$ that contains a configuration $\conf$ with a counter value $\val_\conf>n^2$.
	For $i\leq n$, let $\conf_i$ be the last configuration before $\conf$ and let $\conf'_i$ the first one after $\conf$ in $\run$ such that $\val_{\conf_i}=\val_{\conf'_i}=i$.
	Since $\conf_{\init}$ with $\val_\init=0$ occurs before $\conf$ and $\conf_{\final}$ with $\val_\final=0$ occurs after $\conf$, these configurations are well defined.
	Since there are more such pairs of configurations than pairs of possible states and register assignments, there are two pairs of configurations $(\conf_i, \conf_i)$ and $(\conf_{i+j},\conf'_{i+j})$ that have the same state and register assignments.
	The run has the following form:
	$\run = \conf_{\init} \step[*] \conf_i \step[*] \conf_{i+j} \step[\run'] \conf'_{i+j} \step[*] \conf'_{i} \step[*] \conf_\final $.
	Since $(\conf_{i+j},\conf'_{i+j})$ are defined as the closest configurations to $\conf$ with counter value $i+j$, all configurations in between must have a higher counter value.
	This means we can shorten the run to $\run = \conf_{\init} \step[*] \conf_i \step[\run'] \conf'_{i} \step[*] \conf_\final  $. This is a contradiction to $\run$ being shortest.

	We construct $\RMof{\noadt}$ as follows. We store the counter value in a binary encoding using enough additional registers as bits so that we can encode the number $n^2$.
	If the value becomes higher than $n^2$ our construction blocks $\inc$ operations.
	Since $n^2$ is exponential in the size of $\RMof{\counter}$, the number of registers is polynomial in the size of the input. This can simulate any shortest run that reaches a configuration with $\state_\final$ and $\val_\final=0$ and is thus reachability equivalent with the the original $\rmreachof{\counter}$.

	We know that \rmreachof{\noadt} is in \pspace \cite{par_atig}.
	It follows from the polynomial reduction that \rmreachof{\counter} is in \pspace.
	Applying \autoref{thm:cmplx_pivot_register} gives us that \tsopreachof{\counter} is in \pspace.
\end{proof}

\section{Reachability in TSO($\stack$)}
\label{app:stack}

A \emph{regular language} is a language that can be recognized by a finite state automaton (FSA). A \emph{context-free} language is a language that can be recognized by a pushdown automaton (PDA). We consider the intersection of a set of $n$ regular languages given by the automata $\automaton_1, \dots, \automaton_n$ and one context-free language given by a pushdown automaton $\pushdown$. It is known that deciding emptiness for this intersection is $\exptime$-complete \cite{Hue}. In the following we describe a register machine $\RMof\stack$ with access to a stack, that solves the problem.

The register machine uses registers $\reg_\state^i$ to store the current state of each FSA $\automaton_i$ and another register $\reg_\state^\pushdown$ to store the current state of the PDA $\pushdown$. We assume that these registers are initialized with the initial state of their respective automaton. Furthermore, the register $\reg_\alphabet$ stores the last input symbol that was read. Formally, we can assume that for each automaton, the states are consecutive natural numbers, and the same holds for the input alphabet. This yields a register domain of $\valset = \set{1, \dots, k}$ where $k$ is the maximum size of all automaton state sets and the input alphabet of the PDA. The stack alphabet $\stackalphabet$ of the register machine is the same as the stack alphabet of the PDA $\pushdown$.

Letter by letter, the register machine $\RMof\stack$ non-deterministically guesses an input word $\word \in \alphabet\kstar$ and checks if all automatons can read the current letter.
This is achieved in the following way.
Initially, $\RMof\stack$ is in the state $\state_\init$.
It then chooses a random input symbol $\letter$ and writes it to $\reg_\alphabet$.
Next, $\RMof\stack$ checks if $\pushdown$ can read the current letter:
Let $\tuple{ \state_1, \letter, \stacksymbol, \state_2, \stackword}$ be a transition of $\pushdown$, where in state $\state_1$ the letter $\letter$ is read with $\stacksymbol$ as the topmost stack symbol, and the machine moves to $\state_2$ and writes $\stackword$ onto the stack.
For each such transition in $\pushdown$, there is a sequence of transitions in $\RMof\stack$ that reads $\state_1$ from $\reg_\state^\pushdown$ and $\letter$ from $\reg_\alphabet$, then tries to read $\stacksymbol$ from its stack and writes $\stackword$ onto it.
Lastly, it writes $\state_2$ into $\reg_\state^\pushdown$ and then proceeds to check the first FSA.
The procedure for each FSA is similar, but of course the stack can be ignored.
After all FSAs were successfully moved to their new state, the register machine returns into the state $\state_\init$ and can now proceed with the next letter.
The transition simulation needs a constant number of $\RMof\stack$ states and transitions for each automaton state and transition.

Furthermore, whenever $\RMof\stack$ is in $\state_\init$, it can choose to stop reading input symbols. The machine then checks if all registers $\reg_\state^1, \dots, \reg_\state^n, \reg_\state^\pushdown$ contain accepting states. If that is the case, it moves to the target state $\state_\target$.

In total, the size of the $\RMof\stack$ construction is linear in the size of the input automata.

\begin{lemma}
\label{lem:language_intersection}
	The emptiness of the language intersection of a PDA $\pushdown$ and $n$ FSAs $\automaton_1, \dots, \automaton_n$ can be reduced to the state reachability problem of a register machine $\RMof\stack$ with access to a stack ADT.
\end{lemma}
\begin{proof}
	Following the construction above, the register machine $\RMof\stack$ is in the state $\state_\init$, if and only if there is some word $\word \in \alphabet\kstar$ such that each automaton $\pushdown, \automaton_1, \dots, \automaton_n$ can read the word $w$, starting from their respective initial states.
	Furthermore, the registers $\reg_\state^\pushdown, \reg_\state^1, \dots, \reg_\state^n$ then contain the states of $\pushdown, \automaton_1, \dots, \automaton_n$, respectively, after reading $w$, and the stack of $\RMof\stack$ is equal to the stack of $\pushdown$ after reading $w$.

	This means that if and only if the language intersection is not empty, the register machine $\RMof\stack$ can reach a configuration where it is in state $\state_\init$ and the registers $\reg_\state^\pushdown, \reg_\state^1, \dots, \reg_\state^n$ all contain an accepting state of their respective automaton.
	Then and only then, $\RMof\stack$ can move to the state $\state_\target$.
\end{proof}

\begin{proof}[of \autoref{thm:reach_stack}]
	We show that $\rmreachof\stack$ is $\exptime$-complete.
	It is known that the language intersection problem is $\exptime$-complete \cite{Hue}.
	By \autoref{lem:language_intersection}, it can be reduced to $\rmreachof\stack$.
	Furthermore, the reduction is of polynomial size.
	Thus, $\rmreachof\stack$ is $\exptime$-hard.

	For membership, we encode the registers of $\RMof\stack$ in the states, which yields a finite state machine with access to a stack, i.e. a pushdown automaton. The construction has exponential state overhead. From \cite{Abo}, we have that checking the emptiness of a context-free language generated by a pushdown automaton is polynomial in terms of the size of the automaton. Combined, we get that state reachability of the constructed pushdown automaton is in $\exptime$.

	In summary, $\rmreachof\stack$ is both $\exptime$-hard and in $\exptime$, i.e. it is $\exptime$-complete.
\end{proof}

\section{Well Structured ADTs}
\label{app:wsts}
We briefly review the definitions of well structured transition systems; for details we refer the reader to \cite{Abdulla2000AlgorithmicAO}.

\paragraph{Well structured transitions systems:}
Let $\preceq$ be a relation on the set $\confset$, we say that the transition system given by the tuple $\tuple{\confset, \confset_\init, \lblset, \to}$ is \emph{monotone} w.r.t. $\preceq$ if for every $\conf_1, \conf_2, \conf_3 \in \confset$ we have that $\conf_1 \preceq \conf_2$ and $\conf_1 \to \conf_3$ imply that there exists a run $\run$ such that $\conf_2 \to[\run] \conf_4$ and $\conf_3 \preceq \conf_4$.
Note that this property holds even in the case where $\conf_1 \to[*] \conf_3$, i.e. when there is a run from $\conf_1$ to $\conf_3$.
The relation $\preceq$ is a \emph{well quasi ordering} if for every infinite sequence of configurations $\conf_0, \conf_1, \conf_2, \hdots$ there exist $i, j \in \mathbb{N}$ such that $i < j$ and $\conf_i \preceq \conf_j$.
For $\lbl \in \lblset$, $\conf \in \confset$ and $\confset' \subseteq \confset$, we define
$\conf \upwardclosure = \{ \conf' \in \confset \mid \conf \preceq \conf' \}$,
$\pre_\lbl\of{\confset'} = \{ \conf' \in \confset \mid \conf' \to[\lbl] \conf, \conf \in \confset' \}$,
and $\min\of{\confset'} = \{ \conf \in \confset' \mid \mathop\nexists \conf' \in \confset' : \conf \preceq \conf'\}$.
A relation $\mathop\preceq \subseteq \confset \times \confset$ is decidable, if for any $\conf_1, \conf_2 \in \confset$, we can compute whether $\conf_1 \preceq \conf_2$ holds.

Given a decidable well quasi ordering $\preceq$, we say that the transition system is \emph{well structured}, if it is monotone and the (finite) set $\min(\pre_\lbl(\conf \upwardclosure))$ is computable for every $\conf \in \confset$ and $\lbl \in \lblset$.

\paragraph{Well structured ADTs:}
Given an ADT $\adt = \tuple{ \vals, \set{\val_\init}, \ops, \step\adt }$, let $\preceq \subseteq \vals \times \vals$ be a relation on the set of values $\vals$.
We say that ADT is \emph{monotone} w.r.t. the relation $\preceq$ if for every $\val_1, \val_2, \val_3 \in \vals$ and $\op \in \ops$ we have that if $\val_1 \preceq \val_2$ and $\val_1 \step[\op]\adt \val_3$, then there exists a value $\val_4 \in \vals$ such that $\val_2 \step[\op]\adt \val_4$ and $\val_3 \preceq \val_4$.
Note that this is a stronger condition than for WSTS, where we would only require $\val_2 \step\adt\kstar \val_4$.
An ADT is well structured if it is monotone for a decidable WQO and the $\pre \min$ sets are computable.

\end{document}